\documentclass[aps,prd,twocolumn,groupedaddress,amssymb,eqsecnum,showpacs,
nofootinbib]{revtex4}
\usepackage{graphicx}
\usepackage{bm}

\begin{document}

\preprint{CMU-HEP-03-04}

\title{The fate of the $\alpha$-vacuum}

\author{Hael Collins}
\email{hael@cmuhep2.phys.cmu.edu}
\author{R.~Holman}
\email{rh4a@andrew.cmu.edu}
\author{Matthew R.~Martin}
\email{mmartin@cmu.edu}
\affiliation{Department of Physics, Carnegie Mellon University, 
Pittsburgh PA\ \ 15213}

\date{\today}

\begin{abstract}
de Sitter space-time has a one complex parameter family of invariant vacua for
the theory of a free, massive scalar field.  For most of these vacua, in an
interacting scalar theory the one loop corrections diverge linearly for large
values of the loop momentum.  These divergences are not of a form that can be
removed by a de Sitter invariant counterterm, except in the case of the
Euclidean, or Bunch-Davies, vacuum. 
\end{abstract}

\pacs{04.62.+v,11.10.Gh,98.80.Cq,98.80.Qc}

\maketitle

\section{Introduction}\label{introduction}

The importance of understanding quantum field theory in de Sitter space, the
space-time associated with a positive cosmological constant, has been
heightened by recent observations of both the early and late universe.   The
dramatic results of Wilkinson Microwave Anisotropy Probe \cite{wmap} have
provided further strong evidence that the universe underwent a rapid
inflationary expansion.  Both the large-angle anti-correlation in the
temperature-polarization cross-power spectrum and the nearly flat spectral
index are consistent with the predictions of inflation.  More surprisingly,
the dimming of the type Ia supernovae seen by the Supernova Search Team
\cite{sst} and the Supernova Cosmology Project \cite{scp}, combined with other
observations, is yielding a new standard picture for the contents of the
universe, the largest component of which is a dark energy whose properties are
consistent with a positive cosmological constant.

A striking difference between de Sitter and flat space-time is the richer
vacuum structure of the former.  For a free scalar field in a Minkowski space,
there exists an essentially unique Poincar\'e invariant vacuum state.  In
contrast, for a de Sitter background, Mottola \cite{mottola} and Allen
\cite{allen} discovered an infinite family of vacua for the quantum theory of
a free massive scalar field that are invariant under the isometries  of de
Sitter space.   These vacua can be parameterized by a single complex number,
$\alpha$, and are usually called the $\alpha$-vacua.  Most of these
$\alpha$-vacua have a host of peculiar features,  such as a mixture of
positive and negative frequency modes at short-distances and a non-thermal
behavior that violates the principle of detailed balance.  Only one of these
states, the Euclidean or Bunch-Davies \cite{bunch} vacuum, behaves thermally
when viewed by an Unruh detector \cite{birrelldavies} and reduces to the
Minkowski vacuum as we take the cosmological constant to zero.  The assumption
that the universe was at least approximately in the Euclidean vacuum underlies
the successful predictions of inflation for the calculation of the density
fluctuations which produced the temperature anisotropies in the cosmic
microwave background radiation. 

Despite their unappealing features, the $\alpha$-vacua are perfectly valid
vacua for a free scalar field.  If they cannot be shown to be unphysical, then
their existence would undercut some of the robustness of the inflationary
paradigm---we would need to explain how the epoch prior to inflation managed
to place the universe in the Euclidean vacuum rather than one of the other
infinite family of $\alpha$-vacua.  For example, the regularization needed by
the energy-momentum tensor even for the free theory in the $\alpha$-vacuum is
not generally compatible with that needed after inflation \cite{kklss}.

A complication in formulating quantum field theory in de Sitter space is its
lack of a well-defined $S$-matrix.  In an interacting theory we have two
sources of time dependence for matrix elements---one induced by any inherent
time dependence of the background geometry and another introduced by the
interactions.  In such a system, it is therefore appropriate only to ask time
dependent questions---to study how a matrix element evolves from a given
initial state.  

Schwinger \cite{schwinger} and Keldysh \cite{keldysh} developed a formalism to
solve for this finite time evolution.  In their approach, we specify the state
of the system at an initial time and then evolve to a finite time later.
Here, both the `in' and `out' states correspond to the same state and are
evolved together when we evaluate the expectation value of an operator---in
effect this formalism evaluates matrix elements between two `in' states.  The
Schwinger-Keldysh formalism is thus ideally suited for studying the behavior
of the $\alpha$-vacua in the presence of interactions.  We place the system
initially in an $\alpha$-vacuum and then study whether a sensible evolution
results.  Since the quantum field theory only is evolved over a finite
interval, our results are relevant not only for the more formal question of
the $\alpha$-vacua in an eternal de Sitter background but also for the
phenomenological problem of a finite epoch of inflation.  

The methods established here can also be applied to any initial state, such as
the `truncated $\alpha$-vacua' of \cite{ulf}.  In these vacua, the
short-distance behavior of the $\alpha$-vacua is modified either in accord
with some specific theory, such as the stringy uncertainty relation of
\cite{kempf,gary}, or simply by truncating the $\alpha$ mode functions above
some energy scale to reflect our ignorance of the new physics
\cite{ulf,ss,garygeneric,lowe,brandenberger}.  We address the formal case of a
pure $\alpha$-vacuum and shall study the truncated case later in
\cite{truncated}.

In this article, we show that an interacting scalar field theory in a general
$\alpha$-vacua contains linear divergences which cannot be removed with a de
Sitter invariant renormalization prescription.  These divergences appear in
the one loop corrections and are present for arbitrarily weak interactions.
The specific example we study is the expectation value of the number of
Euclidean particles in an $\alpha$-vacuum.  The divergences appear in the high
momentum region of the loop integral.  We show that they only vanish for the
Euclidean vacuum, which is completely renormalizable.

The subject of the $\alpha$-vacua for an interacting theory has also been
recently investigated in related work \cite{banks,einhorn}.  Both of these
works essentially studied the corrections to the two-point correlation
function obtained between an `in' $\alpha$ state and an `out' state given by
the $\alpha$ state at a later time.  Banks and Mannelli \cite{banks} found
that the interacting theory in the $\alpha$-vacuum required non-local
counterterms while Einhorn and Larsen \cite{einhorn} found pinched
singularities in the loop corrections.  These features provided highly
suggestive evidence that the $\alpha$-vacua are pathological in the presence
of interactions.  Some attempts to modify the theory to avoid these problems
appear in \cite{squeezed,goldstein}. 

We begin with a review of the de Sitter invariant vacua for a free scalar
field in Sec.~\ref{greens}.  This section also shows the form of the Wightman
functions in conformally flat coordinates.  Section \ref{SK} derives the
expectation value of an operator in an interacting theory based on the
Schwinger-Keldysh formalism.  In Sec.~\ref{number} we calculate the change in
the number of Euclidean particles in an $\alpha$-vacuum due to a cubic
interaction and show that in the presence of this interaction, the expectation
value is renormalizable for the Euclidean vacuum while an unrenormalizable
divergence appears for the $\alpha$-vacuum.  Section \ref{discuss} explores
the origin of these divergences in the $\alpha$-vacua in a more general
setting.  We derive the necessary conditions for these divergences to arise
and show how they can appear in a general interacting scalar field theory.
Section \ref{conclude} summarizes our results and suggests future applications
for this formalism.

\section{Green's functions}\label{greens}

In this section we review the rich vacuum structure of a free scalar
field in de Sitter space \cite{bousso}.  We derive the form of the Wightman
function and eventually the Feynman propagator in conformally flat
coordinates.  These Green's functions will be used later for studying the
interacting theory.

The most straightforward method for demonstrating the existence of a family of
de Sitter invariant vacua is to evaluate the two-point Wightman function for a
free massive scalar field in an $\alpha$-vacuum.  For this purpose it is
useful to use a coordinate system that covers the entire space-time.  Such
coordinates are not, however, those best suited for more explicit
calculations.  Therefore, throughout this article we shall study de Sitter
space using conformally flat coordinates,
\begin{equation}
ds^2 = {d\eta^2 - d\vec x^2\over H^2\eta^2} ,
\label{metric}
\end{equation}
with $\eta\in [-\infty,0]$ which cover half of de Sitter space \cite{houches}.
The other half of the space is covered by a set of coordinates with $\eta\to
-\eta$.  These coordinates are simply related to the standard coordinates used
in inflation,
\begin{equation}
ds^2 = dt^2 - e^{2Ht}\, d\vec x^2 , 
\label{inflatemetric}
\end{equation}
through $\eta = - H^{-1}e^{-Ht}$.  $H$ is the Hubble constant and is related
to the cosmological constant by $\Lambda = 6H^2$.

\subsection{The Euclidean vacuum}\label{Euclvac}

To an observer capable only of  probing length scales on which the curvature
of de Sitter space is not apparent, the space-time appears approximately flat.
For the high energy modes then, this observer can apply the same prescription
for defining positive and negative frequency modes as in Minkowski space.  The
vacuum state annihilated by the operators $a_{\vec k}^E$ associated with these
modes corresponds to the Euclidean vacuum.  

The Euclidean vacuum possesses many desirable properties in addition to
matching with the Minkowski vacuum at short distances or as $H\to 0$.  It
corresponds to the unique state whose Wightman function is analytic when
continued to the lower half of the Euclidean sphere.  Moreover, an Unruh
detector placed in the Euclidean vacuum satisfies the principle of detailed
balance as though it were immersed in a thermal system at the de Sitter
temperature, $T_{\rm dS} = H/2\pi$ \cite{bousso}.

If we denote the Euclidean vacuum by $|E\rangle$, the Euclidean Wightman
function for a free massive scalar field $\Phi(x)$ is defined by
\begin{equation}
G_E(x,x') \equiv \langle E | \Phi(x)\Phi(x')| E\rangle .
\label{Ewight}
\end{equation}
Since the metric is spatially flat, the expansion of the scalar field
$\Phi(x)$ in creation and annihilation operators $a_{\vec k}^{E\dagger},
a_{\vec k}^E$ for the vacuum state $|E\rangle$ is  
\begin{equation}
\Phi(\eta,\vec x) = \int {d^3\vec k\over (2\pi)^3}\, \left[ 
U_k^E(\eta) e^{i\vec k\cdot \vec x}\, a^E_{\vec k} 
+ U_k^{E*}(\eta) e^{-i\vec k\cdot \vec x}\, a_{\vec k}^{E\dagger} \right] .
\label{phimodes}
\end{equation}
With the commutator normalized to be 
\begin{equation}
[ a^E_{\vec k}, a_{\vec k'}^{E\dagger} ] 
= (2\pi)^3 \delta^3(\vec k - \vec k') , 
\label{aadagger}
\end{equation}
the Euclidean Wightman function in position space is 
\begin{eqnarray}
G_E(x,x') & = & \int {d^3\vec k\over (2\pi)^3}\, 
e^{i\vec k\cdot (\vec x-\vec x')}\, U_k^E(\eta) U_k^{E*}(\eta')\nonumber \\
& \equiv & \int {d^3\vec k\over (2\pi)^3}\, 
e^{i\vec k\cdot (\vec x-\vec x')}\, G_k^E(\eta,\eta'),
\label{Ewightmodes}
\end{eqnarray}
where the momentum representation the Wightman function is
\begin{equation}
G_k^E(\eta,\eta') = U_k^E(\eta) U_k^{E*}(\eta') . 
\label{ftwightmodes}
\end{equation}
Note that the mode functions only depend on the magnitude of the spatial
momentum, $k=|\vec k|$.

A free massive scalar field satisfies the Klein-Gordon equation,
\begin{equation}
\left[ \nabla^2 + m^2 \right] \Phi(x) = 0 , 
\label{KG}
\end{equation}
so that the mode functions solve the differential equation,
\begin{equation}
\left[ \eta^2 \partial_\eta^2 - 2\eta \partial_\eta + \eta^2 k^2 + m^2H^{-2}
\right] U_k^E(\eta) 
= 0 . 
\label{KGmodes}
\end{equation}
Note that $m^2$ here represents the effective mass of the theory which
includes any contribution from coupling the field to the curvature, $\Phi^2
R$.  In de Sitter space-time the curvature $R$ is constant so this coupling is
of the same form as a mass term.  The solutions to Eq.~(\ref{KGmodes}) are
linear combinations of Bessel functions, 
\begin{equation}
U_k^E(\eta) = c_k\, \eta^{3/2}\, J_\nu(k\eta) 
+ d_k\, \eta^{3/2}\, Y_{\nu}(k\eta) , 
\label{homopart}
\end{equation}
with 
\begin{equation}
\nu = \sqrt{{\textstyle{9\over 4}}-m^2H^{-2}}. 
\label{nudef}
\end{equation}
We shall assume hereafter that $\nu$ is real.  

The general form for the mode functions is applicable to both the Euclidean
vacuum and the $\alpha$-vacuum.  What distinguishes the former is that as the
Hubble constant is taken to vanish, $H\to 0$, so that de Sitter space becomes
flat, we should recover only the positive frequency mode functions,
$e^{-ikt}$.  In the small $H$ limit, 
\begin{equation}
k\eta \to -{ke^{-Ht}\over H} = -{k\over H} + kt + {\cal O}(H) , 
\label{Htozero}
\end{equation}
the leading time dependence of the modes is $U_k^E(\eta) \propto e^{-ikt}$
when 
\begin{equation}
c_k = N_k \qquad
d_k = -i N_k . 
\label{Nkdef}
\end{equation}
Up to the normalization factor, $N_k$, the Euclidean mode functions are given
by 
\begin{eqnarray}
U_k^E(\eta) 
&=& N_k\, \eta^{3/2}\, \left[ J_\nu(k\eta) - i Y_{\nu}(k\eta) \right]
\nonumber \\
&=& N_k\, \eta^{3/2}\, H_\nu^{(2)}(k\eta) .
\label{Emodesunnorm}
\end{eqnarray}
$H^{(2)}_\nu(k\eta)$ represents a Hankel function.  We shall now choose the
units such that $H=1$.  

The normalization is fixed by the canonical equal-time commutation relation
\begin{equation}
\left[ \Pi(\eta,\vec x), \Phi(\eta,\vec x') \right] = -i \delta^3(\vec x -
\vec x') 
\label{equaltime}
\end{equation}
where the conjugate momentum is 
\begin{equation}
\Pi(\eta,\vec x) = {1\over\eta^2} \partial_\eta \Phi(\eta,\vec x) .  
\label{conjmom}
\end{equation}
The equal-time commutation relation requires that the modes satisfy a
Wronskian condition of the form
\begin{equation}
U_k^E\, \partial_\eta U_k^{E*} - \partial_\eta U_k^E\, U_k^{E*} = i \eta^2 ,
\label{wronk}
\end{equation}
which determines the normalization of the modes to be 
\begin{equation}
N_k = {\sqrt{\pi}\over 2} .
\label{Nkval}
\end{equation}
Therefore, the Euclidean mode functions are given by
\begin{equation}
U_k^E(\eta) = {\sqrt{\pi}\over 2} \eta^{3/2} H_\nu^{(2)}(k\eta) .
\label{Emodes}
\end{equation}

While the de Sitter invariance of the Wightman function is not manifest from
Eq.~(\ref{Emodes}), it is possible to write $G_E(x,x')$ as a function of the
de Sitter invariant distance between its arguments \cite{bousso}.  In
conformally flat coordinates, this invariant distance between $x=(\eta,\vec
x)$ and $x'=(\eta',\vec x')$ is 
\begin{equation}
Z(x,x') = {\eta^2 + \eta^{\prime 2} - |\vec x-\vec x'|^2 \over 2\eta\eta'} . 
\label{dSinvinCflat}
\end{equation}

Although we shall state most of our results in terms of the mode functions for
a general mass, it will be convenient to show the results for a particular
case in which the mode functions simplify substantially.  When $\nu={1\over
2}$, the Hankel function in Eq.~(\ref{Emodes}) is proportional to an
exponential, 
\begin{equation}
U_k^E(\eta)\bigr|_{\nu=1/2} = {i\over\sqrt{2k}} \eta e^{-ik\eta} . 
\label{UEhalfnu}
\end{equation}
This case corresponds to a massless, conformally coupled scalar field for
which the effective mass is $m^2=2$.  The Euclidean Wightman function is then 
\begin{equation}
G_E(x,x') = {\eta\eta'\over 16\pi^3} \int {d^3\vec k\over k}\, 
e^{-ik(\eta-\eta')} e^{i\vec k\cdot (\vec x-\vec x')} 
\label{Ewightnuhalf}
\end{equation}
and is finite provided we choose the appropriate $i\epsilon$ prescription,
\begin{equation}
G_E(x,x') = - {1\over 8\pi^2}
{1\over Z - 1 -  i \epsilon\ {\rm sgn}(\eta-\eta') }  .
\label{EwightnuhalfZ}
\end{equation}
Here the appearance of the invariant distance $Z(x,x')$ establishes the de
Sitter invariance of the vacuum.

\subsection{The $\alpha$-vacua}\label{alphavac}

The choice of the short distance behavior of the mode functions which
determined the relative contributions of the two independent solutions to the
Klein-Gordon equation is not the unique choice which leads to a de Sitter
invariant Wightman function.  Mottola \cite{mottola} and Allen \cite{allen}
observed that the vacuum state $|\alpha\rangle$ annihilated by a Bogolubov
transformation of the Euclidean operators,
\begin{equation}
a_{\vec k}^\alpha = N_\alpha \left[ a_{\vec k}^E - e^{\alpha^*} a_{-\vec
k}^{E\dagger} \right] , 
\label{aalphadef}
\end{equation}
also yields a de Sitter invariant Wightman function, 
\begin{equation}
G_\alpha(x,x') \equiv \langle\alpha | \Phi(x)\Phi(x')| \alpha\rangle .
\label{Awight}
\end{equation}
Here, ${\rm Re}\,\alpha < 0$ and the normalization   
\begin{equation}
N_\alpha = \left( 1 - e^{\alpha+\alpha^*} \right)^{-1/2} 
\label{Nalphadef}
\end{equation}
is chosen to preserve the normalization of the commutation relation in the
$\alpha$-vacua, analogous to Eq.~(\ref{aadagger}).  Note that the Euclidean
vacuum is itself among the $\alpha$-vacua being obtained when $\alpha\to
-\infty$.

In proving that $G_\alpha(x,x')$ only depends on $Z(x,x')$ it is useful to use
a coordinatization that covers the entire de Sitter space-time.  In such
global coordinates, both a point $x$ and its antipode $x_A$ occur in the same
coordinate system \cite{bousso}.  It is then possible to chose Euclidean mode
functions $\phi_n^E(x)$ such that $\phi_n^{E*}(x)=\phi_n^E(x_A)$ so that the
Bogolubov transformation of Eq.~(\ref{aalphadef}) gives
\begin{equation}
\phi_n^\alpha(x) = N_\alpha \left[ \phi_n^E(x) + e^\alpha \phi_n^E(x_A)
\right] . 
\label{globalBog}
\end{equation}
Here $n$ labels the elements of a general basis of mode functions.  In this
form, the de Sitter invariance of the Euclidean Wightman function and the fact
that $Z(x,x'_A) = -Z(x,x')$ together imply that the $\alpha$-Wightman function
only depends on the de Sitter invariant distance between $x$ and $x'$
\cite{mottola,allen}.  While it is helpful to use a coordinate system which
contains the antipode of every point to establish this invariance, for
explicit calculations it is not necessary to use global coordinates.
Equation~(\ref{aalphadef}) relates the mode functions of the $\alpha$-vacuum
to the mode functions of the Euclidean vacuum and their complex
conjugates---we do not need to transform the conjugated mode function into a
function of the antipode once we have established that $G_\alpha(x,x')$ is
invariant.

From the Euclidean mode functions of Eq.~(\ref{Emodes}) we can now construct
the mode functions for the $\alpha$-vacua.  Expanding the scalar field in
terms of $\alpha$ creation and annihilation operators,
\begin{equation}
\Phi(\eta,\vec x) = \int {d^3\vec k\over (2\pi)^3}\, \left[ 
U_k^\alpha(\eta) e^{i\vec k\cdot \vec x}\, a^\alpha_{\vec k} 
+ U_k^{\alpha *}(\eta) e^{-i\vec k\cdot \vec x}\, a_{\vec k}^{\alpha\dagger}
\right] ,
\label{phiAmodes}
\end{equation}
and using Eq.~(\ref{aalphadef}) yields
\begin{equation}
U_k^\alpha(\eta) = N_\alpha \left[ 
U_k^E(\eta) + e^\alpha U_k^{E*}(\eta) \right] 
\label{Amodesshort}
\end{equation}
since the $U_k^E(\eta)$ only depend on the magnitude of $\vec k$.  Thus the
$\alpha$-vacuum modes are 
\begin{equation}
U_k^\alpha(\eta) = N_\alpha {\sqrt{\pi}\over 2} \eta^{3/2} \left[ 
H_\nu^{(2)}(k\eta) + e^\alpha H_\nu^{(1)}(k\eta) \right] .
\label{Amodes}
\end{equation}

Inserting the $\alpha$-mode expansion into Eq.~(\ref{Awight}) yields 
\begin{equation}
G_\alpha(x,x') = \int {d^3\vec k\over (2\pi)^3}\, 
e^{i\vec k\cdot (\vec x-\vec x')}\, U_k^\alpha(\eta) U_k^{\alpha *}(\eta') .
\label{Awightmodes}
\end{equation}
Again, the spatial flatness makes it natural to use a momentum representation 
\begin{equation}
G_k^\alpha(\eta,\eta') = U_k^\alpha(\eta) U_k^{\alpha *}(\eta') . 
\label{ftAwightmodes}
\end{equation}

The additional complexity of the mode functions in the $\alpha$-vacuum means
that it is particularly helpful to have a case in which these functions
simplify.  For a massless, conformally coupled scalar field, 
\begin{equation}
U_k^\alpha(\eta)\Bigr|_{\nu=1/2} = N_\alpha {i\over\sqrt{2k}} \eta \left[ 
e^{-ik\eta} - e^\alpha e^{ik\eta} \right] .
\label{Amodesfornuhalf}
\end{equation}
and the Wightman function becomes
\begin{eqnarray}
G_\alpha(x,x') &=& - {N_\alpha^2\over 8\pi^2} \biggl\{
{1\over Z - 1 - i\epsilon\, {\rm sgn}(\eta-\eta') }  \nonumber \\
&&\qquad 
+ {e^{\alpha+\alpha^*}\over Z - 1 + i\epsilon\, {\rm sgn}(\eta-\eta') }
\nonumber \\
&&\qquad 
- {e^\alpha\over Z + 1 - i\epsilon} 
- {e^{\alpha^*}\over Z + 1 + i\epsilon} \biggr\} . \qquad
\label{AwightnuhalfZ}
\end{eqnarray}
As in the Euclidean case, the de Sitter invariance is manifest in the above
expression.

\subsection{Propagation}\label{propa}

To study the propagation of signals in a de Sitter background, define the
Feynman propagator,
\begin{equation}
-iG(x,x') \equiv \langle\alpha |T(\Phi(x)\Phi(x'))| \alpha\rangle ,
\label{Feyndef}
\end{equation}
so that it satisfies the Klein-Gordon equation with a point source, 
\begin{equation}
\left[ \nabla^2_x + m^2 \right] G(x,x') = {\delta^4(x-x')\over\sqrt{-g(x)}} . 
\label{KGFeyn}
\end{equation}
The propagator can only depend on the difference between the spatial positions
of the points so that its Fourier transform is
\begin{equation}
G_k(\eta,\eta') = \int d^3\vec x\, e^{-i\vec k\cdot (\vec x-\vec x')}
G(\eta,\vec x;\eta',\vec x') 
\label{ftGreens}
\end{equation}
where $G_k(\eta,\eta')$ is the solution to 
\begin{eqnarray}
&&\left[ \eta^2 \partial_\eta^2 - 2\eta \partial_\eta + \eta^2 k^2 + m^2
\right] G_k(\eta,\eta') \nonumber \\
&&\qquad\qquad\qquad\qquad\qquad\qquad 
= \eta^2 \eta^{\prime 2} \delta(\eta-\eta') . \qquad
\label{ftGreenseqn}
\end{eqnarray}
The solution to this equation which satisfies the correct boundary conditions
at $\eta'=\eta$ has the form
\begin{equation}
G_k(\eta,\eta') 
= G^>_k(\eta,\eta')\, \Theta(\eta-\eta') 
+ G^<_k(\eta,\eta')\, \Theta(\eta'-\eta) .
\label{Gkpmdef}
\end{equation}
Here $G^>_k(\eta,\eta')$ and $G^<_k(\eta,\eta')$ are essentially the two-point
Wightman functions calculated earlier in Eq.~(\ref{ftAwightmodes}), 
\begin{eqnarray}
G^>_k(\eta,\eta') 
&=& i\, G_k^\alpha(\eta,\eta') 
= i\, U_k^\alpha(\eta)U_k^{\alpha *}(\eta') 
\nonumber \\
G^<_k(\eta,\eta') 
&=& i\, G_k^\alpha(\eta',\eta) 
= i\, U_k^{\alpha *}(\eta)U_k^\alpha(\eta') . \qquad
\label{Gkpmexpand}
\end{eqnarray}

Although propagation for a free field theory in the $\alpha$-vacua contains
some peculiar features, it is not otherwise ill-defined.  The pathological
features of quantum field theory in an $\alpha$-vacuum only appear in an
interacting field theory.  The form of the $\alpha$-Wightman function already
suggests that the interacting theory could be ill-defined, since the various
terms in Eq.~(\ref{AwightnuhalfZ}) contain different $i\epsilon$
prescriptions.  This property implies that in a standard approach to
calculating the one loop corrections to the propagator, among the products of
the propagators participating in the loop appear products of poles with the
opposite $i\epsilon$ prescription---pinched singularities \cite{einhorn}.  For
example, in the one-loop correction to the propagator appears a product of the
Green's functions given in Eq.~(\ref{AwightnuhalfZ}).  However, these pinched
singularities do not by themselves prove whether the $\alpha$-vacuum is itself
pathological or whether the standard methods for studying the quantized theory
are inappropriate for a time-evolving background such as de Sitter space.

\section{The Schwinger-Keldysh formalism}\label{SK}

A significant difference between de Sitter space, in most coordinatizations,
and flat space is the explicit time-dependence of the metric.  Unlike a flat
space-time where the generator of time-translations is a Killing vector
globally, de Sitter space-time has no such global time-like Killing vector.
Moreover, in a particular coordinate system---such as inflationary
coordinates---the time derivative may not even generate an isometry locally.
These properties suggest that rather than attempt to define an $S$-matrix
between `in' and `out' states defined at different times, we should apply a
quantization procedure that evolves an entire matrix element over a finite
interval.  It is also useful to be able to evolve a given state forward from a
specified initial time $\eta_0$, rather than to use a state in the asymptotic
past.  We can always take $\eta_0\to -\infty$.

An additional advantage of solving the evolution over finite intervals is that
such an approach more immediately determines whether the $\alpha$-vacuum is
applicable for inflation, which does not require a de Sitter space-time
eternally, but only over a sufficient interval to generate the number of
$e$-foldings needed to explain the flatness and the homogeneity of the
universe.  If the interacting $\alpha$-vacuum shows its pathology even over a
finite interval, then we can exclude the possibility that the universe was in
a pure $\alpha$-state during any epoch of inflation, regardless of the prior
history of the universe.

The closed time contour formalism developed by Schwinger \cite{schwinger},
Keldysh \cite{keldysh} and Mahanthappa \cite{kt} allows us to study the
evolution of a quantum field theory over a finite interval after specifying
the state at an initial surface.  We review here their approach which leads to
an expression for perturbatively evaluating the matrix element of an operator,
which is given at the end of this section in Eq.~(\ref{Omatrix}).

In the interaction picture, the evolution of operators is given by the free
Hamiltonian, $H_0$, while the evolution of states is given by the
interactions, $H_I$.  If we let $\{ |\Psi\rangle \}$ denote a general basis of
states for the theory, then the behavior of the system is completely described
by the density matrix, $\rho(\eta) = \sum_{\Psi,\Psi'} \rho_{\Psi,\Psi'}
|\Psi\rangle \langle\Psi' |$.  Thus, as the density matrix is constructed from
the states, it satisfies a Schr\" odinger equation of the form
\begin{equation}
i {\partial\over\partial\eta} \rho(\eta) = \bigl[ H_I, \rho(\eta) \bigr] . 
\label{heisenberg}
\end{equation}

The advantage of the interaction picture is that fields evolve using the free
Hamiltonian, 
\begin{equation}
-i {\partial\over\partial\eta} \Phi(\eta,\vec x) = \bigl[ H_0, \Phi(\eta,\vec
x) \bigr] . 
\label{heisenbergPhi}
\end{equation}
The time evolution of the field is precisely that given in the previous
section since here the mode functions still are solutions to the free
Klein-Gordon equation.  

To study the evolution introduced by the interactions, it is convenient to
include a `turning on' function in the interacting part of the Hamiltonian,
\begin{equation}
H = H_0 + \omega(\eta-\eta_0) H_I ; 
\label{hams}
\end{equation}
here $\omega(\eta-\eta_0)$ vanishes when $\eta<\eta_0$ and becomes one when
$\eta$ is sufficiently large compared with $\eta_0$.  Later we shall let this
function be a $\Theta$ step function.  We shall often not write this function
explicitly, absorbing it into $H_I$.  Thus the state does not evolve before
$\eta_0$:  $\rho(\eta) = \rho(\eta_0) \equiv \rho_0$ for $\eta<\eta_0$.

Once we have specified the state at a particular time, $\rho(\eta_0)$, then
the Eq.~(\ref{heisenberg}) allows us to determine the state at all subsequent
times.  To study the vacuum structure of de Sitter space, the initial state
will correspond to an $\alpha$-vacuum.  If we introduce a unitary,
time-evolution operator $U_I(\eta,\eta')$ that evolves the state, 
\begin{equation}
\rho(\eta) = U_I(\eta,\eta_0) \rho(\eta_0) U_I^{-1}(\eta,\eta_0) ,
\label{Uevolve}
\end{equation}
then from Eq.~(\ref{heisenberg}) $U(\eta,\eta_0)$ obeys
\begin{equation}
i {\partial\over\partial\eta} U_I(\eta,\eta_0) = H_I U_I(\eta,\eta_0) 
\label{Ueqn}
\end{equation}
with $U_I(\eta_0,\eta_0)=1$.  The formal solution to this equation is given by
Dyson's equation in terms of the time-ordered exponential
\begin{equation}
U_I(\eta,\eta_0) = T e^{-i\int_{\eta_0}^\eta d\eta^{\prime\prime}\,
H_I(\eta^{\prime\prime})} .
\label{dyson}
\end{equation}

\begin{figure}[!tbp]
\includegraphics{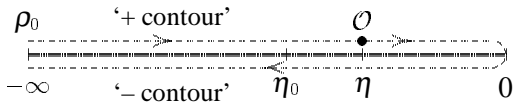}
\caption{The contour used to evaluate the evolution of operators over a finite
time interval.  The initial state is an eigenstate of the Hamiltonian until
$\eta_0$ at which time the interactions are turned on.  We double the field
content so that separate copies of the fields are used for the upper and lower
parts of the contour.\label{ctp}}
\end{figure}

The evolution of the expectation value of an operator in this time-dependent
background is given by 
\begin{eqnarray}
\langle {\cal O} \rangle (\eta) &=& {{\rm Tr} \bigl[ \rho(\eta) {\cal O}
\bigr] \over {\rm Tr} \bigl[ \rho(\eta) \bigr] } \nonumber \\
&=& {{\rm Tr} 
\bigl[ \rho_0 U_I^{-1}(\eta,\eta_0) {\cal O} U_I(\eta,\eta_0) \bigr] \over
{\rm Tr} \bigl[ \rho_0 \bigr] } 
\label{Oevolve}
\end{eqnarray}
Since the state $\rho(\eta_0)$ does not evolve before the interactions are
turned on, we can insert the identity in the form
$U_I(\eta_0,\eta_p)U_I(\eta_p,\eta_0)$ with $\eta_p<\eta_0$ and commute one of
these evolution operators with $\rho_0$ to obtain
\begin{equation}
\langle {\cal O} \rangle (\eta) = {{\rm Tr} 
\bigl[ \rho_0 U_I(\eta_p,\eta) {\cal O} U_I(\eta, \eta_p)\bigr] \over {\rm Tr}
\bigl[ \rho_0 \bigr] } 
\label{OevolveII}
\end{equation}
Inserting another factor of the identity,
$U_I(\eta_p,\eta_f)U_I(\eta_f,\eta_p)$, with $\eta_f>\eta$ yields 
\begin{equation}
\langle {\cal O} \rangle (\eta) = {{\rm Tr} 
\bigl[ \rho_0 U_I(\eta_p,\eta_f) U_I(\eta_f,\eta) {\cal O} U_I(\eta,
\eta_p)\bigr] \over {\rm Tr} \bigl[ \rho_0 \bigr] } 
\label{OevolveIII}
\end{equation}
and finally we let $\eta_p\to -\infty$ and $\eta_f\to 0$, which represent the
infinite past and infinite future in conformal coordinates, so that 
\begin{equation}
\langle {\cal O} \rangle (\eta) = {{\rm Tr} 
\bigl[ U_I(-\infty,0) U_I(0,\eta) {\cal O} U_I(\eta, -\infty) \rho_0 \bigr]
\over {\rm Tr} \bigl[ U_I(-\infty,0) U_I(0, -\infty) \rho_0 \bigr] } 
\label{OevolveIV}
\end{equation}
Reading the operators from right to left in the numerator of this equation,
$\rho_0$ sets the initial state of the system which is then evolved along a
time contour from $-\infty$ to $0$ with an operator inserted at $\eta$; the
final operator evolves back from $0$ to $-\infty$.  The closed time contour
which results is depicted in Fig.~\ref{ctp}.  

To evaluate Eq.~(\ref{OevolveIV}) it is useful to group the evolution
operators into a single time-ordered exponential.  This is accomplished by
formally doubling the field content of the theory, with a set of `$+$' fields
on the increasing-time contour and a set of `$-$' fields on the
decreasing-time contour.  The arrows on the contour indicate time ordering of
events so that events on the $-$ contour always occur after those on the $+$
contour.  We can group the effects of both parts of the contour together by
writing the interacting part of the action appearing in Dyson's equation, as
\begin{equation}
S_I = -\int_{-\infty}^0 d\eta\, H_I(\Phi^+) - \int_0^{-\infty} d\eta\,
H_I(\Phi^-) . 
\label{doubleaction}
\end{equation}
Since the two terms differ only in the direction of the integral over the
conformal time, we can write the action as a single Lagrange density,
\begin{equation}
S_I = - \int_{-\infty}^0 d\eta\, \left[ H_I(\Phi^+) - H_I(\Phi^-) \right] . 
\label{action}
\end{equation}

The field doubling induced by the closed contour effectively doubles the
number of vertices we must include when studying any process---one set with
fields on the $+$ branch and one with fields on the $-$ branch.  From the
Eq.~(\ref{action}), the latter will have couplings with the opposite sign.  In
evaluating matrix elements, Wick contractions produce four propagators for the
possible contractions of pairs of the two types of fields,
\begin{eqnarray}
&&\langle\alpha | T(\Phi^\pm(\eta,\vec x) \Phi^\pm(\eta',\vec x')) |
\alpha\rangle  \nonumber \\
&&\qquad\qquad
= -i \int {d^3\vec k\over (2\pi)^3}\, e^{i\vec k\cdot(\vec x-\vec x')} 
G_k^{\pm\pm}(\eta,\eta') . \qquad
\label{Gofkpmdef}
\end{eqnarray}
The time-ordering of the contractions is determined by the direction along the
contour, 
\begin{eqnarray}
G^{++}_k(\eta,\eta') &=& 
G^>_k(\eta,\eta') \Theta(\eta-\eta') + G^<_k(\eta,\eta') \Theta(\eta'-\eta)
\nonumber \\
G^{--}_k(\eta,\eta') &=& 
G^>_k(\eta,\eta') \Theta(\eta'-\eta) + G^<_k(\eta,\eta') \Theta(\eta-\eta')
\nonumber \\
G^{-+}_k(\eta,\eta') &=& G^>_k(\eta,\eta') \nonumber \\
G^{+-}_k(\eta,\eta') &=& G^<_k(\eta,\eta') , 
\label{Greensmp}
\end{eqnarray}
with the Wightman functions given in Eq.~(\ref{Gkpmexpand}).

Assembling the ingredients of the Schwinger-Keldysh formalism---the general
expression for an operator expectation value in Eq.~(\ref{OevolveIV}), Dyson's
equation (\ref{dyson}) and Eq.~(\ref{action})---provides an explicit
expression for the evolution of $\langle{\cal O}\rangle(\eta)$.  If we let the
initial density matrix be that for a pure $\alpha$-vacuum, then
Eq.~(\ref{OevolveIV}) becomes 
\begin{equation}
\langle\alpha | {\cal O} | \alpha\rangle(\eta) = 
{\langle\alpha | T \left\{ {\cal O}_I e^{-i\int_{-\infty}^0 d\eta\,
[H_I(\Phi^+) - H_I(\Phi^-)] } \right\} | \alpha\rangle \over
\langle\alpha | T \left\{ e^{-i\int_{-\infty}^0 d\eta\, [H_I(\Phi^+) -
H_I(\Phi^-)] } \right\} | \alpha\rangle } .
\label{Omatrix}
\end{equation}
Here we have absorbed any `turning on' function in $H_I$---in essence the time
integrals begin at $\eta_0$.  The time ordering has allowed us to group the
time evolution operators in Eq.~(\ref{OevolveIV}) along the two contours into
a single operator.  This equation for the finite evolution of the expectation
value of ${\cal O}$ is the analogue of the standard $S$-matrix expression used
in Minkowski space.  The virtue of the field doubling is that it removes any
acausal behavior from the matrix element since the $\Theta$-functions in the
propagators combine to limit the upper end of the conformal time integrals to
$\eta$ \cite{schwinger,keldysh}.  This property will become clear in the
calculation of a specific example.

\section{Evolution of the number operator}\label{number}

We now evaluate the expectation value of the number of Euclidean particles in
the $\alpha$-vacuum using the Schwinger-Keldysh formalism \cite{dan}.  This
number operator provides a good measure of whether a particular choice for the
vacuum state becomes pathological in the presence of interactions.  From the
perspective of the Euclidean vacuum, the $\alpha$-vacuum is an excited state.
We can determine the stability of this state when the interactions are turned
on by following the time evolution of the change in the Euclidean number
operator evaluated in the $\alpha$-vacuum.

In the non-interacting theory, we should not encounter any infinite Euclidean
particle production in the $\alpha$-vacuum.  When the interactions are turned
on, some further particle production will occur in the Euclidean vacuum, but
the rate per unit volume should be finite and any divergences which appear
perturbatively must be renormalizable.  What we shall discover is that in the
interacting theory, the $\alpha$-vacuum produces a new class of divergences
that cannot be removed with the usual set of counterterms.  

These divergences in the $\alpha$-vacuum occur at each time in the integrand
as we propagate from some initial state at $\eta_0$ to a finite time later so
that the theory diverges even for an arbitrarily short time after the
interactions are turned on.  This behavior indicates that the $\alpha$-vacuum
is inappropriate even for a finite inflationary epoch.

Although we examine in detail the evolution of the Euclidean number operator
in this section, all of the divergences we find are generic to the one loop
corrections to an arbitrary operator.  Furthermore, while we consider a scalar
theory with a cubic interaction since it has a simple, non-trivial self-energy
correction, similar loop integrals occur in any interacting scalar field
theory.  These more general cases are treated in the next section.  

We first show how the free Hamiltonian can produce a non-trivial, but finite,
time dependence.  We then set up the calculation for the expectation value of
the derivative of the number operator, expressing it in terms of the scalar
field and its conjugate momentum.  The corrections to this expectation value
to leading, non-trivial order in the coupling are then calculated and we
obtain general expressions for the one-loop corrections and the counterterms. 

In the interaction picture, the Hamiltonian is divided into free and
interacting parts,
\begin{equation}
H = H_0 + \Theta(\eta-\eta_0)\, H_I .
\label{hamiltonian}
\end{equation}
Here we have included a $\Theta$-function so that before $\eta_0$ the system
evolves freely; we are always free to take $\eta_0\to -\infty$.  For
conformally flat de Sitter coordinates, the free Hamiltonian for a scalar
field $\Phi(\eta,\vec x)$ of mass $m$ is given by 
\begin{equation}
H_0 = \int d^3\vec x\, \Bigl[ 
{\textstyle{1\over 2}} \eta^2 \Pi^2 
+ {\textstyle{1\over 2}} \eta^{-2} (\vec\nabla\Phi)^2
+ {\textstyle{1\over 2}} \eta^{-4} m^2 \Phi^2
\Bigr] . 
\label{freeham}
\end{equation}
An important difference between de Sitter space and flat space is that the
free Hamiltonian is not diagonal in terms of creation and annihilation
operators, 
\begin{eqnarray}
H_0 &=& {1\over 2} {1\over\eta^2} \int {d^3\vec k\over (2\pi)^3}\, \biggl\{
\left[ a_{\vec k}^E a_{\vec k}^{E\dagger} 
+ a_{\vec k}^{E\dagger} a_{\vec k}^E \right] g_k^E(\eta)
\nonumber \\
&&\qquad\qquad
+ a_{\vec k}^E a_{-\vec k}^E f_k^E(\eta) 
+ a_{\vec k}^{E\dagger} a_{-\vec k}^{E\dagger} f_k^{E*}(\eta)
\biggr\} \qquad
\label{freeHamilton}
\end{eqnarray}
where 
\begin{eqnarray}
g_k^E &=& \partial_\eta U_k^E \partial_\eta U_k^{E*} 
+ \left[ k^2 + {m^2\over\eta^2} \right] U_k^E U_k^{E*}
\nonumber \\
f_k^E &=& \partial_\eta U_k^E \partial_\eta U_k^E 
+ \left[ k^2 + {m^2\over\eta^2} \right] U_k^E U_k^E . 
\label{gkfkA}
\end{eqnarray}
Note that this property holds also for the $\alpha$-vacuum, $E\to\alpha$.  

The origin of the off-diagonal terms in the free Hamiltonian lies in the fact
that the surfaces of constant conformal time are not orthogonal to the
generator of an isometry.  This effect introduces an additional non-trivial
source of time evolution which combines with that produced by the evolution of
the state $|\alpha\rangle$ when the fields interact.  For a time derivative of
a generic operator, we formally have 
\begin{eqnarray}
\partial_\eta \langle {\cal O} (\eta) \rangle 
&=& {\rm Tr} \bigl[ (\partial_\eta\rho) {\cal O} + \rho\partial_\eta {\cal O}
\bigr] \nonumber \\
&=& i\, {\rm Tr}\, \bigl[ - [H_I, \rho] {\cal O} + \rho [H_0, {\cal O}] \bigr]
\nonumber \\
&=& i\, {\rm Tr}\, \bigl[ \rho [H, {\cal O}] \bigr] = i \langle [H, {\cal O}]
\rangle .
\label{Odotdef}
\end{eqnarray}

The off-diagonal terms in the free Hamiltonian induce an evolution in the
number operator even in the free theory.  Since we wish to explore the effect
of interactions on the $\alpha$-vacuum, we construct the number operator from
creation and annihilation operators, $\tilde a_{\vec k}^\dagger$ and $\tilde
a_{\vec k}$, which satisfy 
\begin{equation}
\tilde a_{\vec k}(\eta_0) = a_{\vec k}^E
\label{Atinitial}
\end{equation}
at the moment the interactions are turned on.  The subsequent evolution in the
interaction picture is given by the solution to 
\begin{equation}
-i {\partial\over d\eta} \tilde a_{\vec k} = \bigl[ H_0, \tilde a_{\vec k}
\bigr] .
\label{aHeisen}
\end{equation}
The solution to this equation can be formally expressed as 
\begin{equation}
a_{\vec k}(\eta) = U_0^{-1}(\eta,\eta_0) a_{\vec k}^E U_0(\eta,\eta_0) 
\label{genAevolve}
\end{equation}
where $U_0(\eta,\eta_0)$ is the time evolution operator for the free part of
the theory, 
\begin{equation}
U_0(\eta,\eta_0) = T e^{-i\int_{\eta_0}^\eta d\eta^{\prime\prime}\,
H_0(\eta^{\prime\prime})} . 
\label{freeDyson}
\end{equation}
From the form for the free Hamiltonian, a general solution to
Eq.~(\ref{aHeisen}) is given by a Bogolubov transformation of the time
independent creation and annihilation operators, 
\begin{equation}
a_{\vec k}^E = \alpha_k(\eta)\, \tilde a_{\vec k}
+ \beta_k(\eta)\, \tilde a_{-\vec k}^\dagger , 
\label{bogol}
\end{equation}
where the coefficients satisfy
\begin{eqnarray}
i\eta^2 \partial_\eta \alpha^*_k &=& g_k^E \alpha_k^* + f_k^E \beta_k 
\nonumber \\
-i\eta^2 \partial_\eta \beta_k &=& g_k^E \beta_k + f_k^{E*} \alpha_k^* . 
\label{bogolcoef}
\end{eqnarray}
The standard normalization of the commutator of the transformed creation and
annihilation operators also requires that
\begin{equation}
|\alpha_k(\eta)|^2 - |\beta_k(\eta)|^2 = 1 . 
\label{bogolnorm}
\end{equation}

The general solution to the coefficient equations~(\ref{bogolcoef}) is of the
form
\begin{eqnarray}
\alpha_k^*(\eta) &=& a_1 U_k^E(\eta) + {a_2\over\eta^2} \partial_\eta
U_k^E(\eta) 
\nonumber \\
\beta_k(\eta) &=& - a_1 U_k^{E*}(\eta) - {a_2\over\eta^2} \partial_\eta
U_k^{E*}(\eta) 
\label{coeffsoln}
\end{eqnarray}
where the constants $a_1$ and $a_2$ should satisfy
\begin{equation}
a_1a_2^* - a_1^*a_2 = -i
\label{normsoln}
\end{equation}
from Eq.~(\ref{bogolnorm}).  If we would like the number operator to count the
number of Euclidean particles at the moment we turn on the interactions,
$\eta=\eta_0$, then 
\begin{equation}
a_1 = -i {\partial_\eta U_k^{E*}(\eta_0)\over\eta_0^2} 
\qquad
a_2 = i U_k^{E*}(\eta_0) . 
\label{initialconds}
\end{equation} 

The number operator constructed from the transformed creation and annihilation
operators, $\tilde a_{\vec k}^\dagger\tilde a_{\vec k}$, has the correct
evolution for an interaction picture operator.  Using Eq.~(\ref{Odotdef}),
even the free Hamiltonian induces some evolution in this number operator, 
\begin{equation}
\bigl[ H_0, \tilde a_{\vec k}^\dagger \tilde a_{\vec k} \bigr]
= {i\over\eta^4} D_{ab} \int d^3\vec x d^3\vec y\, e^{i\vec k\cdot (\vec
x-\vec y)} \Phi(\eta_a,\vec x) \Phi(\eta_b,\vec y) ,
\label{HOcommute}
\end{equation}
where $D_{ab}(\eta)$ is the differential operator,
\begin{equation}
D_{ab}(\eta) \equiv p_1(\eta)\, \partial_{\eta_a} \partial_{\eta_b}
- p_2(\eta)\, (\partial_{\eta_a} + \partial_{\eta_b}) + p_3(\eta) .
\label{D0def}
\end{equation}
Here $\eta_{a,b}$ are only labels to indicate the functions on which the
derivatives act; in the end, $\eta_{a,b}$ are set equal to $\eta$.  The
functions $p_i(\eta)$ are 
\begin{eqnarray}
p_1(\eta) 
&=& {\eta^2\over\eta_0^2}\, \partial_\eta |U_k^E(\eta_0)|^2 , 
\nonumber \\
p_2(\eta) 
&=& {\eta^4\over\eta_0^4}\, |\partial_\eta U_k^E(\eta_0)|^2 
- \left[ k^2 + {m^2\over\eta^2} \right] | U_k^E(\eta_0) |^2 
\nonumber \\
p_3(\eta) 
&=& - {\eta^2\over\eta_0^2}\, \left[ k^2 + {m^2\over\eta^2} \right]
\partial_\eta |U_k^E(\eta_0)|^2 . 
\label{pthree}
\end{eqnarray}

Evaluating Eq.~(\ref{HOcommute}) in the $\alpha$-vacuum gives the tree level
evolution of the number operator in this state, 
\begin{equation}
i \langle\alpha | \bigl[ H_0, \tilde a_{\vec k}^\dagger \tilde a_{\vec k}
\bigr] | \alpha\rangle 
= - {1\over\eta^4} V D_{ab}(\eta) U_k^\alpha(\eta_a) U_k^{\alpha *}(\eta_b) 
\label{Ntree}
\end{equation}
$V$ is the spatial volume.  We can divide the volume from both sides to yield
the particle production rate per unit volume for which this tree contribution
is completely finite, if non-zero.

Now consider a cubic interaction with its associated counterterms,
\begin{equation}
H_I = \int d^3\vec x\, \eta^{-4} \left[ J \Phi 
+ {\textstyle{1\over 2}} \delta m^2 \Phi^2 
+ {\textstyle{1\over 3!}}\lambda \Phi^3 \right] .
\label{Hintdef}
\end{equation}
The cubic vertex will generally introduce one-loop corrections to the
two-point functions, so we have included a mass counterterm, $\delta m^2$.
Since the $\Phi^3$ interaction breaks the $\Phi\to -\Phi$ symmetry of the free
theory, we also expect the interaction to generate graphs containing tadpole
insertions which are cancelled with the correct choice for $J$.  To the order
we shall study, no wave function renormalization is needed.

The change in the number operator induced by these interactions is given by
applying Eq.~(\ref{Omatrix}),
\begin{eqnarray}
&&\!\!\!\!\!\!\!\!\!\!\!\!
\dot N_{\alpha, \vec k}^E(\eta) \equiv 
\label{matrix} \\
&&
{i\langle\alpha | T \left\{ [ H, \tilde a_{\vec k}^\dagger \tilde a_{\vec k} ]
e^{-i\int_{\eta_0}^0 d\eta\, [H_I(\Phi^+) - H_I(\Phi^-)] } \right\} |
\alpha\rangle \over
\langle\alpha | T \left\{ e^{-i\int_{\eta_0}^0 d\eta\, [H_I(\Phi^+) -
H_I(\Phi^-)] } \right\} | \alpha\rangle } .
\nonumber
\end{eqnarray}
The evolution of the free field is simple in the interaction picture, so it is
useful to write the creation and annihilation operators in terms of the field
and its conjugate momentum, by expanding in the time independent operators,
Eq.~(\ref{bogol}), and inverting the operator expansion in
Eq.~(\ref{phimodes}),
\begin{eqnarray}
a^E_{\vec k} &\! = \!& i \int d^3\vec x\, e^{-i\vec k\cdot \vec x} \left[
U_k^{E*}\, \Pi(\eta,\vec x) - \eta^{-2} \partial_\eta U_k^{E*} \Phi(\eta,\vec
x) \right] \nonumber \\
a_{\vec k}^{E\dagger} &\! = \!& - i \int d^3\vec x\, e^{i\vec k\cdot \vec x}
\left[
U_k^E\, \Pi(\eta,\vec x) - \eta^{-2} \partial_\eta U_k^E \Phi(\eta,\vec x)
\right] . \nonumber \\
&&  \label{invert}
\end{eqnarray}
The commutator with the free part of the Hamiltonian is then as was given in
Eq.~(\ref{HOcommute})
while that with the interacting part is
\begin{widetext}
\begin{eqnarray}
\bigl[ H_I, \tilde a_{\vec k}^\dagger \tilde a_{\vec k} \bigr] 
&=& 
{i\over\eta^4} J\, (2\pi)^3 \delta^3(\vec k) \int d^3\vec x\,
\Bigl[ 2 |U_k^E(\eta_0)|^2 \Pi(\eta,\vec x) 
- \eta_0^{-2} \partial_\eta |U_k^E(\eta_0)|^2 \Phi(\eta,\vec x) \Bigr]
\label{HIcommute} \\
&& + {i\over\eta^4} \delta m^2 \int d^3\vec x d^3\vec y\, e^{i\vec k\cdot
(\vec x-\vec y)} 
\Bigl[ |U_k^E(\eta_0)|^2 
\left[ \Phi(\eta,\vec x) \Pi(\eta,\vec y) +  \Pi(\eta,\vec x) \Phi(\eta,\vec
y) \right]
\nonumber \\
&&\qquad\qquad\qquad 
- \eta_0^{-2} \partial_\eta |U_k^E(\eta_0)|^2 
\Phi(\eta,\vec x) \Phi(\eta,\vec y) \Bigr] 
\nonumber \\
&&
+ {i\over\eta^4} {\lambda\over 2} \int d^3\vec x d^3\vec y\, e^{i\vec k\cdot
(\vec x-\vec y)} 
\Bigl[ |U_k^E(\eta_0)|^2 
\left[ \Phi^2(\eta,\vec x) \Pi(\eta,\vec y) +  \Pi(\eta,\vec x)
\Phi^2(\eta,\vec y) \right]
\nonumber \\
&&\qquad\qquad\qquad 
- \eta_0^{-2} U_k^E(\eta_0) \partial_\eta U_k^{E*}(\eta_0) \Phi(\eta,\vec x)
\Phi^2(\eta,\vec y) 
- \eta_0^{-2}  U_k^{E*}(\eta_0) \partial_\eta U_k^E(\eta_0) \Phi^2(\eta,\vec
x) \Phi(\eta,\vec y) \Bigr] . 
\nonumber 
\end{eqnarray}
\end{widetext}
The expectation values of each of these commutators will be evaluated in the
$\alpha$-vacuum to order $\lambda^2$.

\begin{figure}[!tbp]
\includegraphics{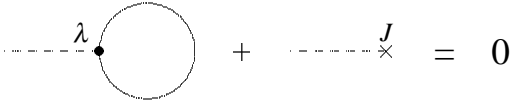}
\caption{The coefficient $J$ of the linear counterterm is chosen to cancel
insertions of tadpoles.  The dashed line represents a line in a general
diagram.
\label{tadpole}}
\end{figure}

Before evaluating the expectation values of these commutators perturbatively,
a large class of graphs, those containing a tadpole subgraph, are eliminated
through the proper choice of the coefficient $J$ of the linear counterterm.
To leading order in $\lambda$, this choice for $J$ is 
\begin{equation}
J = - {\lambda\over 16\pi^3} \int d^3\vec p\, \left|
U_p^\alpha(\eta^{\prime\prime}) \right|^2 . 
\label{Jtadpole}
\end{equation}
This cancellation is shown diagrammatically in Fig.~\ref{tadpole}.  Note that
while the loop integral in Eq.~(\ref{Jtadpole}) contains an apparent time
dependence, it is in fact time independent,
\begin{eqnarray}
&&\int d^3\vec p\, \left| U_p^\alpha(\eta^{\prime\prime}) \right|^2 
\nonumber \\
&&\qquad
= N_\alpha^2 \pi^2 
\int_0^\infty \xi^2\, d\xi\, 
\left| H_\nu^{(2)}(\xi) + e^\alpha H_\nu^{(1)}(\xi) \right|^2 .
\qquad
\label{notimeloop}
\end{eqnarray}

\begin{figure}[!tbp]
\includegraphics{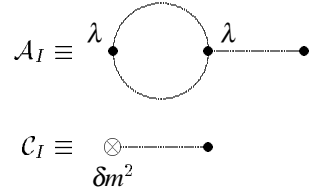}
\caption{After choosing the linear term to cancel graphs containing a tadpole,
only these terms contribute at order $\lambda^2$ to the expectation value of
$[H_I,\tilde a_{\vec k}^\dagger \tilde a_{\vec k}]$ in the $\alpha$-vacuum.
The first represents a self-energy graph while the second is from the mass
counterterm.
\label{loopI}}
\end{figure}

The form of the leading corrections to the expectation value of $[H_I, \tilde
a_{\vec k}^\dagger \tilde a_{\vec k}]$ in the $\alpha$-vacuum are simpler
since the commutator is already itself of order $\lambda$.  The order
$\lambda^2$ corrections to this commutator are shown diagrammatically in
Fig.~\ref{loopI}.  The other corrections from the cubic interaction at this
order contain tadpole subgraphs and are cancelled when Eq.~(\ref{Jtadpole}) is
satisfied.  

One of the subtleties in evaluating the evolution of the number operator is
that it contains time derivatives.  For example, the $\delta m^2$ term of
Fig.~\ref{loopI} contains a term of the form 
\begin{equation}
\langle\alpha | T \left( \Phi(\eta,\vec x) \Pi(\eta,\vec y) +
\Pi(\eta,\vec x) \Phi(\eta,\vec x) \right) | \alpha\rangle 
\label{timePi}
\end{equation}
which can produce Schwinger terms if the time ordering of the operators is not
treated carefully.  A method for avoiding such terms, following \cite{dan}, is
to write the canonical momentum as 
\begin{equation}
\Pi(\eta,\vec x) = \lim_{\eta^{\prime\prime}\to\eta} 
{1\over\eta^{\prime\prime 2}} 
\partial_{\eta^{\prime\prime}} \Phi(\eta^{\prime\prime},\vec x)
\label{schwing}
\end{equation}
and then to place the fields on the appropriate contours so that the time
ordering naturally given along the contour preserves the correct ordering of
the operators in Eq.~(\ref{timePi}), 
\begin{eqnarray}
&&\lim_{\eta^{\prime\prime}\to\eta} {1\over\eta^{\prime\prime 2}}
\partial_{\eta^{\prime\prime}}
\langle\alpha | T \bigl( \Phi^-(\eta,\vec x) 
\Phi^+(\eta^{\prime\prime},\vec y) + \nonumber \\
&&\qquad\qquad\qquad\qquad
\Phi^-(\eta^{\prime\prime},\vec x) \Phi^+(\eta,\vec x) \bigr) | \alpha\rangle
. \qquad
\label{timePipm}
\end{eqnarray}
With this prescription, the mass counterterm in Fig.~\ref{loopI} contributes
\begin{eqnarray}
{\cal C}_I &=& 
- {\delta m^2\over\eta^6} V \biggl\{ 
|U_k^E(\eta_0)|^2\, \partial_\eta |U_k^\alpha(\eta)|^2 
\nonumber \\
&&\qquad\quad
- {\eta^2\over\eta_0^2} \partial_\eta |U_k^E(\eta_0)|^2\, 
|U_k^\alpha(\eta)|^2 \biggr\}
\quad
\label{Cone}
\end{eqnarray}
at lowest order.  

The only non-trivial effect of the interactions on expectation value of the
commutator with the interaction Hamiltonian arises from the self-energy graph,
\begin{eqnarray}
{\cal A}_I &\! =\! & 
- {\lambda^2\over\eta^6} {V\over 4\pi^3} |U_k^E(\eta_0)|^2
\nonumber \\
&&\quad\times
\int_{\eta_0}^\eta 
{d\eta'\over\eta^{\prime 4}}\, {\rm Im}\bigl[
\partial_\eta U_k^\alpha(\eta) U_k^{\alpha *}(\eta') L_k^\alpha(\eta,\eta')
\bigr]
\nonumber \\
&& 
+ {\lambda^2\over\eta^4} {V\over 8\pi^3} 
{\partial_\eta |U_k^E(\eta_0)|^2\over\eta_0^2} 
\nonumber \\
&&\quad\times
\int_{\eta_0}^\eta 
{d\eta'\over\eta^{\prime 4}}\, {\rm Im} \bigl[
U_k^\alpha(\eta) U_k^{\alpha *}(\eta') L_k^\alpha(\eta,\eta') \bigr]
\qquad 
\label{Aone}
\end{eqnarray}
where the loop integral is given by
\begin{equation}
L_k^\alpha(\eta,\eta') \equiv 
\int d^3\vec p\, U_p^\alpha(\eta) U_p^{\alpha *}(\eta') 
U_{p-k}^\alpha(\eta) U_{p-k}^{\alpha *}(\eta') . 
\label{loopdef}
\end{equation}

\begin{figure}[!tbp]
\includegraphics{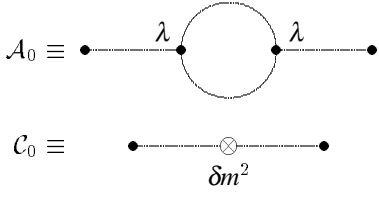}
\caption{The order $\lambda^2$ corrections to the expectation value of
$[H_0,a_{\vec k}^{E\dagger} a_{\vec k}^E]$ in the $\alpha$-vacuum.  Again, the
first represents a self-energy graph while the second is from the mass
counterterm.
\label{loop0}}
\end{figure}

In addition to the corrections in Eq.~(\ref{Cone}) and Eq.~(\ref{Aone}), the
fact that the free Hamiltonian is not the conserved quantity associated with a
time-like generator of an isometry of de Sitter space means that the
expectation value of $[H_0, \tilde a_{\vec k}^\dagger \tilde a_{\vec k}]$ also
contributes at order $\lambda^2$.  The linear counterterm also cancels the
graphs containing a tadpole subgraph and any vacuum to vacuum disconnected
graphs are removed by the denominator of Eq.~(\ref{matrix}).  The only
remaining corrections then are those shown in Fig.~\ref{loop0}.
These graphs contribute 
\begin{eqnarray}
{\cal A}_0 &=& - {\lambda^2\over\eta^4} {V\over 2\pi^3} D_{ab}(\eta) 
\nonumber \\
&&\times
\int_{\eta_0}^\eta {d\eta_1\over\eta_1^4} 
\int_{\eta_0}^{\eta_1} {d\eta_2\over\eta_2^4}\,  
{\rm Im}\bigl[ U_k^\alpha(\eta_a) U_k^{\alpha *}(\eta_1) \bigr]
\nonumber \\
&&\qquad\qquad\times {\rm Im}
\bigl[ U_k^\alpha(\eta_b) U_k^{\alpha *}(\eta_2) L_k^\alpha(\eta_1,\eta_2) 
\bigr] \qquad
\label{Anaught}
\end{eqnarray}
and 
\begin{eqnarray}
{\cal C}_0 &=& - {2\delta m^2\over\eta^4} V D_{ab}(\eta) 
\nonumber \\
&&
\int_{\eta_0}^\eta {d\eta_1\over\eta_1^4}
{\rm Im} \bigl[ U_k^\alpha(\eta_a) U_k^{\alpha *}(\eta_1) 
U_k^\alpha(\eta_b) U_k^{\alpha *}(\eta_1) \bigr]
\qquad
\label{Cnaught}
\end{eqnarray}
to the expectation value of the derivative of the number operator, $\dot
N_{\alpha,\vec k}^E$.  

An important feature to note is that the mass counterterm insertions, in
Eq.~(\ref{Cone}) and Eq.~(\ref{Cnaught}), do not vanish in the Euclidean
limit; they are needed to cancel a logarithmic divergence in the corresponding
self-energy diagrams.

\subsection{Renormalizing the Logarithmic Divergence}

The one loop corrections in the generic $\alpha$-vacuum and in the Euclidean
limit differ in their divergence structure.  Both cases contain a logarithmic
divergence which can be regularized and cancelled by the appropriate choice
for the mass counterterm, $\delta m^2$.  What distinguishes the general
$\alpha$-vacuum from the Euclidean vacuum is the appearance of an additional
class of terms that diverge linearly in the loop momentum.  This divergence
cannot be renormalized.  In this subsection, we summarize the renormalization
of the logarithmic divergence for the case of the massless conformally coupled
scalar field in the Euclidean vacuum.  The detailed calculation for the
$\alpha$-vacuum is left for the Appendix \ref{dimen}.

Formally, the evolution of the Euclidean number operator to order $\lambda^2$
in the Euclidean vacuum is given by 
\begin{eqnarray}
\dot N^E_{E,\vec k}(\eta) 
&=& - {1\over\eta^4} V D_{ab}(\eta) U_k^E(\eta_a) U_k^{E*}(\eta_b)
\nonumber \\
&&+ \lim_{\alpha\to -\infty} \left[ {\cal A}_I + {\cal C}_I
+ {\cal A}_0 + {\cal C}_0 \right]
\label{eucllim}
\end{eqnarray}
The case of a massless, conformally coupled scalar field is most readily
analyzed since the mode functions have a simple form given in
Eq.~(\ref{UEhalfnu}).  The tree contribution in this case is given by
\begin{equation}
{V\over 2k^2\eta^3} \left[ {\eta^4\over\eta_0^4} 
+ k^2\eta^2 {\eta^2\over\eta_0^2}
+ {\eta\over\eta_0} - k^2\eta_0^2 - {\eta_0^2\over\eta^2}
\right]
\label{Ntreenuhalf}
\end{equation}
which is finite for $\eta\in[\eta_0,0)$.  The loop integral for $\nu={1\over
2}$, which occurs in ${\cal A}_I$, is 
\begin{equation}
L^E_{k,\nu={1\over 2}}(\eta,\eta') \equiv 
- {i\pi\over 2} {(\eta\eta')^2\over\eta-\eta'} e^{-ik(\eta-\eta')} ,
\label{Euclloophalf}
\end{equation}
so that 
\begin{eqnarray}
{\cal A}_I 
&=& {\lambda^2\over k^2} {V\over 32\pi^2} 
\left[ {\eta_0^2\over\eta^4} - {1\over\eta_0\eta} \right] 
\int_{\eta_0}^\eta {d\eta'\over\eta-\eta'}\, \cos 2k(\eta-\eta') 
\nonumber \\
&& - {\lambda^2\over k} {V\over 32\pi^2} {\eta_0^2\over\eta^3} 
\int_{\eta_0}^\eta {d\eta'\over\eta-\eta'}\, \sin 2k(\eta-\eta') . 
\label{AoneEnuhalf}
\end{eqnarray}
The corresponding counterterm is 
\begin{equation}
{\cal C}_I 
= - {\delta m^2\over k^2} {V\over 2}
\left[ {\eta_0^2\over\eta^5} - {1\over\eta_0\eta^2} \right] . 
\label{ConeEnuhalf}
\end{equation}
The second integral in Eq.~(\ref{AoneEnuhalf}) is completely finite at all
times $\eta<0$.  The first integral, however, contains a logarithmic
divergence at the upper end of the $d\eta'$ integration.  This term can be
regularized as described in the appendix~\ref{dimen}.  The pole as the
regularization is removed, $\epsilon\to 0$, is given by 
\begin{equation}
{\cal A}_I = {\lambda^2\over k^2} {V\over 32\pi^2} 
\left[ {\eta_0^2\over\eta^5} - {1\over\eta_0\eta^2} \right] 
{1\over\epsilon} 
+ {\rm finite} , 
\label{AoneEnuhalfpole}
\end{equation}
and is cancelled by choosing 
\begin{equation}
\delta m^2 = {1\over\epsilon} {\lambda^2\over 16\pi^2} 
\label{Euclmct} 
\end{equation}
in Eq.~(\ref{ConeEnuhalf}).  The analogous logarithmic divergence in ${\cal
A}_0$ is cancelled by the counterterm ${\cal C}_0$.

The self-energy diagrams in the $\alpha$-vacuum also contain a logarithmic
divergence which can be removed by a suitable choice for the mass counterterm, 
\begin{equation}
\delta m^2 = {1\over\epsilon} {\lambda^2\over 16\pi^2} 
\left( 1 + e^{\alpha+\alpha^*} \right) N_\alpha^2 ,
\label{alphamct}
\end{equation}
which reduces to that for the Euclidean case in Eq.~(\ref{Euclmct}).  The
origin and regularization of this divergence is discussed more fully in
Appendix~\ref{dimen}.

\subsection{The Linear Divergence of the $\alpha$-vacuum}

Including the order $\lambda^2$ corrections, the derivative of the number of
Euclidean particles in the $\alpha$ vacuum is again given by the sum of the
contributions shown in Fig.~\ref{loopI} and Fig.~\ref{loop0} as well as the
tree level contribution of Eq.~(\ref{Ntree}), 
\begin{eqnarray}
\dot N^E_{\alpha,\vec k}(\eta) &=&
- {1\over\eta^4} V D_{ab}(\eta) U_k^\alpha(\eta_a) U_k^{\alpha *}(\eta_b) 
\nonumber \\
&& + {\cal A}_I + {\cal C}_I + {\cal A}_0 + {\cal C}_0 . 
\label{alphanum}
\end{eqnarray}
Each self-energy graph contains a loop integral, Eq.~(\ref{loopdef}).  Unlike
the Euclidean case which is completely finite once we have established an
$i\epsilon$ prescription, the loop integral over $\alpha$-mode functions
diverges linearly in the spatial momentum.  Introducing a bound $\Lambda$ to
remove the large momenta in the loop, $\int_0^\infty p^2\, dp \int d\Omega_2
\to \int_0^\Lambda p^2\, dp \int d\Omega_2$, assuming $\Lambda > |\vec k|$,
the divergent part of Eq.~(\ref{loopdef}) is 
\begin{eqnarray}
L_k^\alpha(\eta_1,\eta_2) &=& 
e^{\alpha+\alpha^*} N_\alpha^4 {2\pi\Lambda\over k} (\eta_1\eta_2)^2 
\label{loopdivg} \\
&&\quad\times
\left[ {\sin k(\eta_1-\eta_2)\over\eta_1-\eta_2}
+ {\sin k(\eta_1+\eta_2)\over\eta_1+\eta_2} \right] 
\nonumber \\
&& +\ {\rm finite} . 
\nonumber
\end{eqnarray}
The appearance of the factor $e^{\alpha+\alpha^*}$ shows why such a divergent
term does not arise in the Euclidean limit.  

Unlike the logarithmic divergence, this divergence cannot be removed by a
momentum independent value for $\delta m^2$.  For example, the divergent piece
of the self-energy graph in Fig.~\ref{loopI} is 
\begin{eqnarray}
{\cal A}_I &=&
- e^{\alpha+\alpha^*} N_\alpha^4 {\lambda^2\over k\eta^4} 
{V\Lambda\over 2\pi^2} |U_k^E(\eta_0)|^2 
\label{Aonedivg} \\
&&\int_{\eta_0}^\eta {d\eta'\over\eta^{\prime 2}}\, 
{\rm Im}\, \Bigl\{ \partial_\eta U_k^E(\eta) U_k^{E*}(\eta') \Bigr\} 
\nonumber \\
&&\qquad\times
\left[ {\sin k(\eta-\eta')\over\eta-\eta'}
+ {\sin k(\eta+\eta')\over\eta+\eta'} \right] 
\nonumber \\
&&
+ e^{\alpha+\alpha^*} N_\alpha^4 {\lambda^2\over k\eta^2} 
{V\Lambda\over 4\pi^2} {\partial_\eta |U_k^E(\eta_0)|^2\over\eta_0^2} 
\nonumber \\
&&\int_{\eta_0}^\eta {d\eta'\over\eta^{\prime 2}}\, 
{\rm Im}\, \Bigl\{ U_k^E(\eta) U_k^{E*}(\eta') \Bigr\} 
\nonumber \\
&&\qquad\times
\left[ {\sin k(\eta-\eta')\over\eta-\eta'}
+ {\sin k(\eta+\eta')\over\eta+\eta'} \right] + \cdots 
\nonumber 
\end{eqnarray}
which cannot be cancelled by a momentum-independent choice for $\delta m^2$ in
Eq.~(\ref{Cone}).

To observe this incompatibility of the momentum dependence of the divergences
in the $\alpha$-vacua loop corrections and the available counterterms, at
least in the case of a massless, conformally coupled scalar field, it is
sufficient to compare the limiting behavior in the external momentum of the
loop corrections and the corresponding counterterms.  In the $k=|\vec k|\to 0$
limit the divergent parts of both ${\cal A}_I$ and ${\cal A}_I$ scale as
$k^{-1}\Lambda$.  In contrast, the leading $k$-dependence of the counterterms
scale as $k^{-2}$, at least when $\alpha$ is real, which is required in any
case for a ${\cal CPT}$ invariant theory \cite{bousso}.  Thus, no choice of
$\delta m^2$, which does not depend on $k$, is possible such that ${\cal C}_I$
and ${\cal C}_0$ cancel the divergence in ${\cal A}_I$ and ${\cal A}_0$ as
$\Lambda\to\infty$.

Note that the linear divergence is not present in the opposite quantity---the
expectation value of the number of $\alpha$-particles in the Euclidean vacuum.
This quantity is given by expressions similar to those above except with the
labels interchanged for the mode functions, $U_q^\alpha \leftrightarrow
U_q^E$, and some of the $\alpha$-dependent coefficients are slightly altered
in Eqs.~(\ref{Cone})--(\ref{Cnaught}).  The crucial difference is that the
presence of the Euclidean states in Eq.~(\ref{Omatrix}) leads to Euclidean
propagators so that in particular the loop integral is over Euclidean modes,
as in Eq.~(\ref{Euclloop}), for which no linear divergence occurs.

\section{Discussion}\label{discuss}

The linear divergence that arises from the one loop corrections in the
$\alpha$-vacuum is a UV effect.  At arbitrarily short distances there exists
an interference of the positive and negative frequency modes which cancels the
rapidly oscillating phases among some of the terms within the loop integral.
Without such a cancellation, these phases could damp these high-momentum
contributions through an appropriate $i\epsilon$ prescription.  This
interference of phases is a specific feature of the propagator in the
$\alpha$-vacuum and does not occur in the Euclidean case.  In this section we
shall discuss the origin of this divergence and determine the necessary
conditions for it to arise.

Consider a loop containing $n$ vertices connected by $n$ internal
propagators---those through which the common loop momentum flows.  Since in de
Sitter space it is convenient to perform a Fourier transform over only the
spatial coordinates, each vertex has a time, $\eta_i$ for $i=1,\ldots, n$,
associated with it.  Eventually, we integrate over all these times as they
arise from the exponent of the time evolution operator in Eq.~(\ref{Omatrix}).
To determine whether a particular loop can produce a UV divergence, we must
first count the powers of momentum in the high loop momentum region.

Let $G_{p-k_i}^>(\eta_i,\eta_{i+1})$ represent the Wightman function within a
loop propagator connecting the $i$ and $(i+1)$ vertices.  The loop momentum is
$\vec p$ and $\vec k_i$ denotes other momenta following through the $i$-th
leg.  In the UV limit $|\vec p-\vec k_i| \gg |\eta_i|^{-1},
|\eta_{i+1}|^{-1}$, the leading behavior of this Wightman function is 
\begin{eqnarray}
&&\!\!\!\!\!\!\!\!
G_{p-k_i}^>(\eta_i,\eta_{i+1}) \to
\label{highpofWight} \\
&& \!\!\!\!\!\!
iN_\alpha^2 {\eta_i\eta_{i+1}\over 2|\vec p-\vec k_i|} 
\Bigl[ 
e^{-i|\vec p-\vec k_i|(\eta_i-\eta_{i+1})} 
+ e^{\alpha+\alpha^*} e^{i|\vec p-\vec k_i|(\eta_i-\eta_{i+1})} 
\nonumber \\
&&\!\!\!\!\!\!
- i e^\alpha e^{-i\pi\nu} e^{i|\vec p-\vec k_i|(\eta_i+\eta_{i+1})} 
+ i e^{\alpha^*} e^{i\pi\nu} e^{-i|\vec p-\vec k_i|(\eta_i+\eta_{i+1})} 
\Bigr] . 
\nonumber
\end{eqnarray}
Note that the propagator contains factors of both
$G_{p-k_i}^>(\eta_i,\eta_{i+1})$ and $G_{p-k_i}^<(\eta_i,\eta_{i+1}) =
G_{p-k_i}^>(\eta_{i+1},\eta_i)$.  For the purpose of the power counting, it is
important to note that, aside from the phases, in the UV limit
$G_{p-k_i}^\alpha(\eta_i,\eta_{i+1}) \sim p^{-1}$.  Thus in integrating over a
loop, 
\begin{equation}
\int^\Lambda d^3\vec p\, \prod_{i=1}^n G_{p-k_i}^\alpha(\eta_i,\eta_{i+1})
\sim 
\int^\Lambda {dp\over p^{n-2}} . 
\end{equation}
We only encounter a possible UV divergence if $n\le 3$.  Note that the $n=1$
case can be removed by a counterterm since the loop only depends on the loop
momentum and not on any other momenta in the graph.

\begin{figure}[!tbp]
\includegraphics{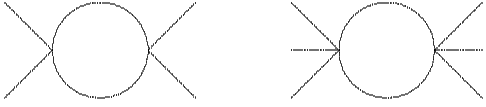}
\caption{Examples of divergent diagrams in theories with quartic (left) or
quintic (right) interactions.  Any loop that contains only two lines will
exhibit a UV divergence similar to that in Eq.~(\ref{loopdivg}).  This result
follows from the structure of the propagator in the $\alpha$-vacuum and not
the form of the interactions.
\label{loop4}}
\end{figure}

The $n=2$ case can produce a linear divergence.  Since the divergence only
depends on the form of the propagator and not the form of the interaction,
such divergences generically occur in any interacting theory, for example in
the processes shown in Fig.~\ref{loop0} and Fig.~\ref{loop4}.
Superficially, a divergence might seem possible even in the Euclidean case if
a product of Wightman functions, $G_p^>(\eta_1,\eta_2)
G_{p-k}^<(\eta_1,\eta_2)$, occurs in the loop integral.  However, the
Schwinger-Keldysh formalism is constructed to remove such terms and only
factors of $G_p^>(\eta_1,\eta_2) G_{p-k}^>(\eta_1,\eta_2)$, and
$G_p^<(\eta_1,\eta_2) G_{p-k}^<(\eta_1,\eta_2)$ can appear in the loop
integral in which the $p$-dependent phases do not cancel in the UV.  The
important difference in the $\alpha$-vacuum is that each Wightman function
contains terms whose $p$-dependent phases have the opposite signs.  Thus even
after the Schwinger-Keldysh formalism has been applied, a product of the two
loop Green's functions, 
\begin{eqnarray}
&&\!\!\!\!\!
G_p^>(\eta_1,\eta_2) G_{p-k}^>(\eta_1,\eta_2) \to
\label{highpofWightA} \\
&&\qquad
- e^{\alpha+\alpha^*} N_\alpha^4 {(\eta_1\eta_2)^2\over 4p|\vec p-\vec k|} 
\Bigl[ 
e^{-ip(\eta_1-\eta_2)} e^{i|\vec p-\vec k|(\eta_1-\eta_2)}
\nonumber \\
&&\qquad\qquad\qquad\qquad
+ e^{ip(\eta_1-\eta_2)} e^{-i|\vec p-\vec k|(\eta_1-\eta_2)} 
\nonumber \\
&&\qquad\qquad\qquad\qquad
+ e^{ip(\eta_1+\eta_2)} e^{-i|\vec p-\vec k|(\eta_1+\eta_2)} 
\nonumber \\
&&\qquad\qquad\qquad\qquad
+ e^{-ip(\eta_1+\eta_2)} e^{i|\vec p-\vec k|(\eta_1+\eta_2)} 
\Bigr] + \cdots
\nonumber
\end{eqnarray}
will have some phase cancellation as $p\to\infty$.  What renders these terms
unrenormalizable, however, is that although the $p$-dependent phases cancel in
the divergent terms, they still contain a non-trivial dependence on the
momentum entering the loop from the rest of the diagram.  

This analysis of the phase structure of the two-propagator loop does not
necessarily show that such divergences cannot be removed from the theory
through a suitable renormalization prescription.  This fact can only be
established by summing all the contributions to this graph and demonstrating
that the dependence of the resulting divergent term on the momenta external to
the loop is incompatible with a counterterm insertion, as was done in the
previous section.  However, since all loop integrals containing only two
propagators have essentially the same structure, given by Eq.~(\ref{loopdef}),
we see that the $\alpha$-vacuum cannot be renormalized in any interacting
theory, regardless of the form of the interaction.

\begin{figure}[!tbp]
\includegraphics{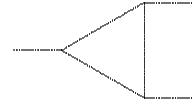}
\caption{Diagrams containing a loop with three propagators can be
logarithmically divergent in the loop momentum.  For example, a vertex
correction in a $\Phi^3$ theory could generate such a divergence.
\label{vertex}}
\end{figure}

The power counting argument indicates that a logarithmic divergence can arise
for a loop with three legs, such as the vertex correction graph shown in
Fig.~\ref{vertex}.  The loop integral contains a product of three terms of the
form of Eq.~(\ref{highpofWight}), or its complex conjugate, which generally
contains terms where the high loop momentum dependence of the phase cancels.  
For example, in Fig.~\ref{vertex} if the coordinates associated with the three
vertices are $(\eta_1,\vec k_1)$, $(\eta_2,\vec k_2)$ and $(\eta_3,-\vec k_1 -
\vec k_2)$, then in the product of three propagators occur terms such as 
\begin{eqnarray}
&&\!\!\!\!\!\!\!\!\!\!\!
G_{p-k_1}^>(\eta_1,\eta_2) G_{p+k_2}^>(\eta_2,\eta_3) G_p^>(\eta_1,\eta_3) \to
\label{logdivofvert} \\
&&
-i e^{\alpha+\alpha^*} N_\alpha^6 
{(\eta_1\eta_2\eta_3)^2\over 8p |\vec p-\vec k_1| |\vec p+\vec k_2|} 
\nonumber \\
&&\times 
e^{-i|\vec p-\vec k_1|(\eta_1-\eta_2)} 
e^{-i|\vec p+\vec k_2|(\eta_2-\eta_3)} 
e^{-ip(\eta_3-\eta_1)} 
+ \cdots .
\nonumber
\end{eqnarray}
In the high momentum region of the loop momentum, the phase factor will be
independent of the integrated momentum and the integral will be
logarithmically divergent.  As with the self-energy case before, these
arguments can only demonstrate under what conditions a divergence can occur.
Whether these logarithmic terms cancel among each other or whether the
resulting divergence can be removed by a counterterm requires performing the
full integration and summing all the relevant products of Wightman functions.
However, since we have already seen that the self-energy graphs exhibit a
pathological behavior in the $\alpha$-vacuum, we shall not study these vertex
corrections further here.

\section{Conclusions}\label{conclude}

The preceding discussion shows that a class of linear divergences from loops
with two propagators---and logarithmic divergences from loops with three
propagators---generically appears in any interacting theory in an
$\alpha$-vacuum.  These divergences arise from the form of the $\alpha$-vacuum
propagator, which is determined by the free field Hamiltonian, and not on the
detailed form of the interaction.  What the Schwinger-Keldysh formalism allows
is a precise statement of the problem of an interacting theory in an
$\alpha$-vacuum.  With a high-momentum cut off, we can find all the terms that
diverge linearly with this cut off and analyze their dependence on the momenta
external to the loop.  The resulting expressions are not cancelled by a set of
de Sitter invariant counterterms of the same form as those in the original
Lagrangian.  The appearance of $\alpha$-dependent prefactors also shows why
such terms do not plague the Euclidean vacuum.

The fact that the divergence originates from high momentum modes and the form
of the $\alpha$-dependent prefactors provides the basis of a simple heuristic
explanation for the divergence.\footnote{We would like to thank Dan Boyanovsky
for suggesting this kinetic interpretation}  The number density of Euclidean
particles per unit volume at the time that the interactions are turned on is
\begin{equation} 
n_{\vec k}^\alpha \equiv V^{-1}
\langle\alpha | a_{\vec k}^{E\dagger} a_{\vec k}^E | \alpha\rangle 
= e^{\alpha+\alpha^*} N_\alpha^2 
= {1\over e^{-\alpha-\alpha^*} - 1} 
\label{occupy}
\end{equation}
since $\tilde a_{\vec k}(\eta_0) = a_{\vec k}^E$.  From the perspective of the
Euclidean vacuum, the $\alpha$-vacuum looks like a distribution whose
occupation number is given by $n_{\vec k}^\alpha$.  Note that $n_{\vec
k}^\alpha$ is actually independent of $\vec k$.  If we then replace the
$\alpha$-dependent terms in the propagators with the factors $n_{\vec
k}^\alpha$, maintaining the momentum labels, then among the many divergent
terms contributing to ${\cal A}_I$ occurs the expression,
\begin{eqnarray} 
{\cal A}_I 
&=& - {\lambda^2\over k\eta} {V\over 64\pi^3} 
{\partial_\eta |U_k^E(\eta_0)|^2\over\eta_0^2}\, 
\int_{\eta_0}^\eta {d\eta'\over \eta'} 
\int {d^3\vec p\over p|\vec p-\vec k|}\, 
\nonumber \\
&&\quad\times
\bigl[ 
(n_{\vec k}^\alpha +1)(n_{\vec p}^\alpha +1)n_{\vec p-\vec k}^\alpha
- n_{\vec k}^\alpha n_{\vec p}^\alpha (n_{\vec p-\vec k}^\alpha +1)
\bigr]
\nonumber \\
&&\quad\times
\sin\bigl[[p+k-|\vec p - \vec k|](\eta-\eta') \bigr] \nonumber \\
&&
+ \cdots , 
\label{kinetic}
\end{eqnarray}
in the massless conformally coupled case.  This expression resembles a `gain
minus loss' process in the $\alpha$-background---for example, one part
describes the creation of two particles from one while the other describes the
creation of one from the annihilation of two.  Since $n_{\vec p}^\alpha$ is
constant, nothing suppresses the large $p$ divergence.

This divergence is only present in a true $\alpha$-vacuum and not in a
`truncated $\alpha$-vacuum'---a state that  is set equal to a Euclidean vacuum
above some scale $|\vec k| > M$ \cite{ulf}.  For a truncated $\alpha$-vacuum,
the $n_{\vec p}^\alpha$'s vanish above $M$ so integrals such as
Eq.~(\ref{kinetic}) become finite.  The largest contribution to the change in
the number operator scales as $\lambda^2 M$.  These truncated $\alpha$-vacua
have no divergences, although exactly how the state evolves may depend on how
the truncation is implemented.  Using our formalism, it becomes possible to
study how one of these vacua evolves during inflation and the amount by which
it would alter the appearance of angular power spectrum of the cosmic
microwave background radiation \cite{truncated}.

To conclude then, interactions destabilize the $\alpha$-vacua. We have
calculated a physical quantity, the conformal time rate of change of the
number of Euclidean mode particles in the $\alpha$-vacuum, and found that it
diverges.  This divergence reflects a physical pathology of an interacting
theory in a true $\alpha$-vacuum.

\begin{acknowledgments}
This work was supported in part by DOE grant DE-FG03-91-ER40682.  We are
grateful to Dan Boyanovsky for helpful discussions.  R.~H.~thanks M.~Einhorn,
F.~Larsen and Jim Cline for helpful discussions, as well as the Aspen Center
for Physics where this work was begun.

\end{acknowledgments}

\appendix

\section{Regularization of logarithmic divergences}\label{dimen}

The self-energy diagrams in the Fig.~\ref{loopI} and Fig.~\ref{loop0} contain
two classes of divergences in the $\alpha$-vacuum.  One class diverges
linearly in the magnitude of the spatial loop momentum.  The dependence of
this divergence on the external momentum flowing through the graph is not of
the same form as that appearing from the insertion of one of the available
counterterms.  In this sense, these divergences can not be renormalized and
indicate a pathological feature of an interacting theory in the
$\alpha$-vacuum.  However, these linearly divergent terms vanish in the
Euclidean limit.

The second class of divergences exist in both the Euclidean and the general
$\alpha$-vacua.  These divergences depend logarithmically on the conformal
time and can, at least in the massless conformally coupled case where the
calculation simplifies, be removed by a constant mass counterterm.  It is
important to establish the renormalization of this type of divergence not only
to show that the evolution of the number operator is finite for an interacting
scalar field in the Euclidean vacuum, but also in the $\alpha$-case.  If we
wish to consider a `truncated $\alpha$-vacuum', one which is cut off at some
high energy scale such as the Planck mass \cite{ulf}, we thereby remove the
linear divergence of Eq.~(\ref{loopdivg}), but the logarithmic divergence is
still present in ${\cal A}_I$ and ${\cal A}_0$.  In this appendix, we
demonstrate how to renormalize this divergence.

In the massless conformally coupled case, loop integral over $\alpha$-vacuum
propagators of Eq.~(\ref{loopdef}) yields 
\begin{eqnarray}
&&
\!\!\!\!\!
L^\alpha_{k,\nu={1\over 2}}(\eta_1,\eta_2) = 
\nonumber \\
&&
- {i\pi\over 2} N_\alpha^2 (\eta_1\eta_2)^2 
{e^{-ik(\eta_1-\eta_2)} + e^{\alpha+\alpha^*} e^{ik(\eta_1-\eta_2)} \over
\eta_1-\eta_2} 
\nonumber \\
&& 
+ {i\pi\over 2} N_\alpha^4 (\eta_1\eta_2)^2 
[e^\alpha-e^{\alpha^*}] {e^\alpha e^{ik(\eta_1+\eta_2)} + e^{\alpha^*}
e^{-ik(\eta_1+\eta_2)} \over \eta_1+\eta_2} 
\nonumber \\
&& 
- {i\pi\over k} e^\alpha N_\alpha^4 \eta_1\eta_2 
\sin k\eta_1 \left[ e^{ik\eta_2} - e^{2\alpha^*} e^{-ik\eta_2} \right] 
\nonumber \\
&& 
+ {i\pi\over k} e^{\alpha^*} N_\alpha^4 \eta_1\eta_2 
\sin k\eta_2 \left[ e^{-ik\eta_1} - e^{2\alpha} e^{2ik\eta_1} \right] 
\nonumber \\
&& 
+ {2\pi\over k} (\Lambda - k) e^{\alpha+\alpha^*} N_\alpha^4 (\eta_1\eta_2)^2 
\nonumber \\
&&\qquad\qquad\times
\left[ {\sin k(\eta_1-\eta_2)\over \eta_1-\eta_2} 
+ {\sin k(\eta_1+\eta_2)\over \eta_1+\eta_2} \right] .
\label{loopalpha} 
\end{eqnarray}
The final term is the linearly divergent term.  Among the remaining terms,
only the first produces any logarithmic divergence when integrated over the
conformal time.  In the case of either self-energy contribution,
Eq.~(\ref{Aone}) or Eq.~(\ref{Anaught}), we integrate $\eta_2$ from $\eta_0$
to $\eta_1$ ($\eta_1\to\eta$ for the ${\cal A}_I$ graph) and encounter a
singularity from the $(\eta_1-\eta_2)$ denominator of imaginary part of the
first term in Eq.~(\ref{loopalpha}), 
\begin{eqnarray}
&&
\!\!\!\!\!\!\!\!
L^\alpha_{k,\nu={1\over 2}}(\eta_1,\eta_2) = 
\label{loopalphapole} \\
&&
- {i\pi\over 2} (1+e^{\alpha+\alpha^*}) N_\alpha^2 (\eta_1\eta_2)^2 
{\cos k(\eta_1-\eta_2)\over \eta_1-\eta_2} 
+ \cdots .
\nonumber 
\end{eqnarray}
Inserting this result in the self-energy contributions yields 
\begin{eqnarray}
{\cal A}_I &=&
{\lambda^2\over k\eta^4} {V\over 16\pi^2} 
(1+e^{\alpha+\alpha^*}) N_\alpha^4 |U_k^E(\eta_0)|^2 
\nonumber \\
&&\times
\int_{\eta_0}^\eta {d\eta'\over\eta'(\eta-\eta')}\, 
\cos k(\eta-\eta')
\nonumber \\ 
&&\quad \times\biggl\{
(1+e^{\alpha+\alpha^*}) \cos k(\eta-\eta')
\nonumber \\ 
&&\quad\qquad 
- \left( e^\alpha e^{ik(\eta+\eta')} + e^{\alpha^*}e^{-ik(\eta+\eta')}
\right) 
\nonumber \\ 
&&\quad\qquad 
-ik\eta \left( e^\alpha e^{ik(\eta+\eta')} - e^{\alpha^*}e^{-ik(\eta+\eta')} 
\right) 
\biggr\}
\nonumber \\ 
&&
- {\lambda^2\over k\eta} {V\over 32\pi^2} 
(1+e^{\alpha+\alpha^*}) N_\alpha^4 
{\partial_\eta |U_k^E(\eta_0)|^2\over\eta_0^2} 
\nonumber \\
&&\times
\int_{\eta_0}^\eta {d\eta'\over\eta'(\eta-\eta')}\, 
\cos k(\eta-\eta')
\nonumber \\ 
&&\quad \times\biggl\{
(1+e^{\alpha+\alpha^*}) \cos k(\eta-\eta')
\nonumber \\ 
&&\quad\qquad 
- \left( e^\alpha e^{ik(\eta+\eta')} + e^{\alpha^*}e^{-ik(\eta+\eta')}
\right) 
\biggr\}
\nonumber \\ 
&&+ \cdots 
\label{logofAone}
\end{eqnarray}
and
\begin{eqnarray}
{\cal A}_0 &=& 
- {\lambda^2\over k^2\eta^4} {V\over 16\pi^2} (1+e^{\alpha+\alpha^*})
N_\alpha^4 D_{ab}(\eta)\, 
\nonumber\\
&&\times
\int_{\eta_0}^\eta {d\eta_1\over\eta_1}
\int_{\eta_0}^{\eta_1} {d\eta_2\over\eta_2}\, 
\eta_a\eta_b {\cos k(\eta_1-\eta_2)\over\eta_1-\eta_2} 
\nonumber \\
&&\qquad\times
\sin k(\eta_a-\eta_1) 
\Bigl[ (1+e^{\alpha+\alpha^*})\cos k(\eta_b-\eta_2) 
\nonumber \\
&&\qquad\qquad\quad
- e^\alpha e^{ik(\eta_b+\eta_2)} - e^{\alpha^*} e^{-ik(\eta_b+\eta_2)}
\Bigr]
\nonumber \\
&&
+ \cdots 
\label{logofAnaught}
\end{eqnarray}
In both of these equations, the ellipses indicate terms which do not diverge
logarithmically in the conformal time.

Both Eq.~(\ref{logofAone}) and Eq.~(\ref{logofAnaught}) contain divergent
integrals of the form
\begin{equation}
\int_{\eta_0}^{\eta_1} {e^{2iq\eta_2}\, d\eta_2\over\eta_2(\eta_1-\eta_2)} ,
\label{logdiverge} 
\end{equation}
where $q$ is $k$, $0$, or $-k$.  Changing to a dimensionless variable, 
\begin{equation}
r \equiv 1 - {\eta_2\over\eta_1} ,
\label{taudef}
\end{equation}
Eq.~(\ref{logdiverge}) becomes 
\begin{equation}
\int_{\eta_0}^{\eta_1} {e^{2iq\eta_2}\, d\eta_2\over\eta_2(\eta_1-\eta_2)} 
= - {e^{2iq\eta_1}\over\eta_1} \int_0^{1 - {\eta_0\over\eta_1} } 
{e^{-2iq\eta_1 r}\, dr\over(r - 1)r} . 
\label{shift} 
\end{equation}
We can regularize this integral by inserting a factor of $r^\epsilon$ in the
integrand and then extract the pole as $\epsilon\to 0$,
\begin{equation}
\int_0^{1 - {\eta_0\over\eta_1} } 
{e^{-2iq\eta_1 r}\, r^{\epsilon-1}\, dr\over r-1} =
- {1\over\epsilon} + {\rm finite} . 
\label{regularize} 
\end{equation}
Thus, 
\begin{equation}
\int_{\eta_0}^{\eta_1} {e^{2iq\eta_2}\, d\eta_2\over\eta_2(\eta_1-\eta_2)} 
= {e^{2iq\eta_1}\over\eta_1} {1\over\epsilon} + {\rm finite} . 
\label{polepiece} 
\end{equation}

Applying this regularization to the self-energy graphs gives 
\begin{eqnarray}
&&\!\!\!\!\!\!\!\!\!\!\!{\cal A}_I =
\nonumber \\
&&
{\lambda^2\over\eta^6} {V\over 16\pi^2} {1\over\epsilon}
(1+e^{\alpha+\alpha^*}) N_\alpha^2 
\nonumber \\
&&\times\biggl[
|U_k^E(\eta_0)|^2 \partial_\eta |U_k^\alpha(\eta)|^2
- {\eta^2\over\eta_0^2}
\partial_\eta |U_k^E(\eta_0)|^2 |U_k^\alpha(\eta)|^2 \biggr]
\nonumber \\ 
&&+ \cdots 
\label{poleofAone}
\end{eqnarray}
and
\begin{eqnarray}
{\cal A}_0 &=& 
{2\lambda^2\over\eta^4} {V\over 16\pi^2} {1\over\epsilon}
(1+e^{\alpha+\alpha^*}) N_\alpha^2 D_{ab}(\eta)\, 
\nonumber\\
&&\times
\int_{\eta_0}^\eta {d\eta_1\over\eta_1^4}\, 
{\rm Im}\left\{ U_k^\alpha(\eta_a) U_k^{\alpha *}(\eta_1) 
U_k^\alpha(\eta_b) U_k^{\alpha *}(\eta_1) 
\right\} 
\nonumber \\
&&
+ \cdots . 
\label{poleofAnaught}
\end{eqnarray}
In deriving Eq.~(\ref{poleofAnaught}) we have used the fact that the operator
$D_{ab}(\eta)$ is symmetric in $\eta_a$ and $\eta_b$.  Both poles are removed
by the appropriate mass counterterm graphs ${\cal C}_I$ and ${\cal C}_0$ given
in Eq.~(\ref{Cone}) and Eq.~(\ref{Cnaught}), respectively, when 
\begin{equation}
\delta m^2 = {1\over\epsilon} {\lambda^2\over 16\pi^2} 
\left( 1 + e^{\alpha+\alpha^*} \right) N_\alpha^2 . 
\label{alphamctA}
\end{equation}


\begin{thebibliography}{99}

\bibitem{wmap}
H.~V.~Peiris {\it et al.},
astro-ph/0302225.

\bibitem{sst}
A.~G.~Riess {\it et al.}  [Supernova Search Team Collaboration],
Astron.\ J.\  {\bf 116}, 1009 (1998) [astro-ph/9805201].

\bibitem{scp}
S.~Perlmutter {\it et al.}  [Supernova Cosmology Project Collaboration],
Astrophys.\ J.\  {\bf 517}, 565 (1999) [astro-ph/9812133].

\bibitem{mottola}
E.~Mottola,
Phys.\ Rev.\ D {\bf 31}, 754 (1985).

\bibitem{allen}
B.~Allen,
Phys.\ Rev.\ D {\bf 32}, 3136 (1985).

\bibitem{bunch}
T.~S.~Bunch and P.~C.~Davies,
Proc.\ Roy.\ Soc.\ Lond.\ A {\bf 360}, 117 (1978).

\bibitem{birrelldavies}
N.~D.~Birrell and P.~C.~W.~Davies, {\it Quantum Fields in Curved Space\/}
(Cambridge University Press, Cambridge, England, 1989).

\bibitem{kklss}
N.~Kaloper, M.~Kleban, A.~Lawrence, S.~Shenker and L.~Susskind,
JHEP {\bf 0211}, 037 (2002) [hep-th/0209231].

\bibitem{schwinger}
J.~S.~Schwinger,
J.\ Math.\ Phys.\  {\bf 2}, 407 (1961).

\bibitem{keldysh}
L.~V.~Keldysh,
Zh.\ Eksp.\ Teor.\ Fiz.\  {\bf 47}, 1515 (1964)
[Sov.\ Phys.\ JETP {\bf 20}, 1018 (1965)].

\bibitem{ulf}
U.~H.~Danielsson,
Phys.\ Rev.\ D {\bf 66}, 023511 (2002) [hep-th/0203198]; 
U.~H.~Danielsson,
JHEP {\bf 0212}, 025 (2002) [hep-th/0210058].

\bibitem{kempf}
A.~Kempf,
Phys.\ Rev.\ D {\bf 63}, 083514 (2001) [astro-ph/0009209]; 
A.~Kempf and J.~C.~Niemeyer,
Phys.\ Rev.\ D {\bf 64}, 103501 (2001) [astro-ph/0103225].

\bibitem{gary}
R.~Easther, B.~R.~Greene, W.~H.~Kinney and G.~Shiu,
Phys.\ Rev.\ D {\bf 64}, 103502 (2001) [hep-th/0104102]; 
R.~Easther, B.~R.~Greene, W.~H.~Kinney and G.~Shiu,
Phys.\ Rev.\ D {\bf 67}, 063508 (2003) [hep-th/0110226].

\bibitem{ss}
S.~Shankaranarayanan,
Class.\ Quant.\ Grav.\  {\bf 20}, 75 (2003) [gr-qc/0203060].

\bibitem{garygeneric}
R.~Easther, B.~R.~Greene, W.~H.~Kinney and G.~Shiu,
Phys.\ Rev.\ D {\bf 66}, 023518 (2002) [hep-th/0204129].

\bibitem{lowe}
K.~Goldstein and D.~A.~Lowe,
Phys.\ Rev.\ D {\bf 67}, 063502 (2003) [hep-th/0208167].

\bibitem{brandenberger}
J.~Martin and R.~Brandenberger,
hep-th/0305161.

\bibitem{truncated}
H.~Collins, R.~Holman and M.~Martin, in preparation.

\bibitem{banks}
T.~Banks and L.~Mannelli,
Phys.\ Rev.\ D {\bf 67}, 065009 (2003)
[hep-th/0209113].

\bibitem{einhorn}
M.~B.~Einhorn and F.~Larsen,
Phys.\ Rev.\ D {\bf 67}, 024001 (2003) [hep-th/0209159].

\bibitem{squeezed}
M.~B.~Einhorn and F.~Larsen,
hep-th/0305056.

\bibitem{goldstein}
K.~Goldstein and D.~A.~Lowe,
hep-th/0302050.

\bibitem{bousso}
R.~Bousso, A.~Maloney and A.~Strominger,
Phys.\ Rev.\ D {\bf 65}, 104039 (2002) [hep-th/0112218].

\bibitem{houches}
For an excellent review of de Sitter space, its causal structure and its
coordinatizations see M.~Spradlin, A.~Strominger and A.~Volovich,
hep-th/0110007.

\bibitem{kt}
K.~T.~Mahanthappa, Phys.\ Rev. {\bf 126}, 329 (1962); P.~M.~Bakshi and
K.~T.~Mahanthappa, J.\ Math.\ Phys.\ {\bf 41}, 12 (1963).

\bibitem{dan}
D.~Boyanovsky, H.~J.~de Vega and S.~Y.~Wang,
Phys.\ Rev.\ D {\bf 61}, 065006 (2000) [hep-ph/9909369].

\end{thebibliography}
\end{document}